\documentclass[fleqn,usenatbib]{mnras}

\usepackage{newtxtext,newtxmath,ulem}

\usepackage[T1]{fontenc}

\newcommand{\tyomaadded}[1]{#1}

\newcommand{\newadded}[1]{#1}

\DeclareRobustCommand{\VAN}[3]{#2}
\let\VANthebibliography\thebibliography
\def\thebibliography{\DeclareRobustCommand{\VAN}[3]{##3}\VANthebibliography}

\usepackage{graphicx}	
\usepackage{amsmath}	
\usepackage{pdflscape}  
\usepackage{animate}    

\title[ASKAP rapid scintillators]{ASKAP observations of multiple rapid scintillators reveal a degrees-long plasma filament}

\author[Y. Wang et al.]{Yuanming Wang,$^{1,2}$\thanks{E-mail: ywan3191@uni.sydney.edu.au}
Artem Tuntsov,$^3$\thanks{E-mail: Artem.Tuntsov@manlyastrophysics.org}
Tara Murphy,$^{1,4}$\thanks{E-mail: tara.murphy@sydney.edu.au}
Emil Lenc,$^2$
Mark Walker,$^3$
\newauthor
Keith Bannister,$^2$
David L. Kaplan,$^5$
Elizabeth K. Mahony$^2$
\\
$^{1}$Sydney Institute for Astrophysics, School of Physics, The University of Sydney, Sydney, NSW 2006, Australia\\
$^{2}$CSIRO Astronomy and Space Science, Australia Telescope National Facility, PO Box 76, Epping, NSW 1710, Australia\\
$^{3}$Manly Astrophysics, 15/41-42 East Esplanade, Manly 2095, Australia\\
$^{4}$ARC Centre of Excellence for Gravitational Wave Discovery (OzGrav), Hawthorn, Victoria, Australia\\
$^{5}$Department of Physics, University of Wisconsin–Milwaukee, Milwaukee, Wisconsin 53201, USA.
}

\date{Accepted 2021 January 13. Received 2021 January 12; in original form 2020 November 17}

\pubyear{2015}

\begin{document}
\label{firstpage}
\pagerange{\pageref{firstpage}--\pageref{lastpage}}
\maketitle

\begin{abstract}
We present the results from an Australian Square Kilometre Array Pathﬁnder search for radio variables on timescales of hours. 
We conducted an untargeted search over a 30 deg$^2$ field, with multiple 10-hour observations separated by days to months, at a central frequency of 945 MHz. 
We discovered six rapid scintillators from 15-minute model-subtracted images with sensitivity of $\sim 200\,\mu$Jy/beam; two of them are extreme intra-hour variables with modulation indices up to $\sim 40\%$ and timescales as short as tens of minutes. 
Five of the variables are in a linear arrangement on the sky with angular width $\sim 1$~arcmin and length $\sim$ 2 degrees, revealing the existence of a huge plasma filament in front of them.
We derived kinematic models of this plasma from the annual modulation of the scintillation rate of our sources, and we estimated its likely physical properties: a distance of $\sim 4$ pc and length of $\sim 0.1$ pc.
The characteristics we observe for the scattering screen are incompatible with published suggestions for the origin of intra-hour variability leading us to propose a new picture in which the underlying phenomenon is a cold tidal stream.
This is the first time that multiple scintillators have been detected behind the same plasma screen, giving direct insight into the geometry of the scattering medium responsible for enhanced scintillation. 
\end{abstract}




\section{Introduction}

Radio sources with angular size $\lesssim 1$ mas (e.g. active galactic nuclei (AGN) or pulsars) can be affected by propagation effects caused by irregularities in the ionised interstellar medium (ISM) of the Milky Way, causing them to show variations with typical timescales from \newadded{minutes} to months \citep[e.g.][]{1972ApL....12..193H, 1984A&A...134..390R}. 
Some flat-spectrum AGN have been found to exhibit \textit{flickering} with short timescales from minutes to few days and larger amplitude fluctuations of up to $\sim 50\%$, known as either intraday variability (IDV) or intrahour variability (IHV) depending on the timescale \citep[e.g.][]{1987AJ.....94.1493H, 1997ApJ...490L...9K}. 
These rapid variables have been confirmed as interstellar scintillation (ISS) due to their arrival time delays through scintillation pattern \citep{2002Natur.415...57D}. 
The annual modulation of timescales, caused by the relative speed between the Earth and the screen varying in a year, is further evidence of scintillation origin \citep{2000aprs.conf..147J, 2001ApJ...550L..11R}. 

Extreme IHV with large amplitude modulations ($\gtrsim 10\%$) requires both the small angular size of the AGN of order microarcseconds, and highly structured, nearby ($\lesssim$ tens of pcs from the solar system) scattering medium in front of the source \citep{2002ApJ...581..103R, 2003A&A...404..113D, 2007MNRAS.380L..20M}. 
The inferred brightness temperatures of the background AGN are usually high ($\gtrsim 10^{12}$ K), suggesting a large Doppler boosting factor $\gtrsim 10$ \citep[e.g.][]{2000ApJ...529L..65D, 2000ApJ...538..623M}, greater than observed in existing Very Long Baseline Interferometry (VLBI) surveys \citep[e.g.][]{2020ApJS..247...57C}. 

The physical nature of the unusual scattering medium remains unknown. 
It is widely accepted that the required pressure fluctuations in the plasma screen are much higher than those in the typical extended ISM \citep{2002ApJ...581..103R, 2017ApJ...843...15W} and hence over-dense scattering mediums may be commonplace in the Galaxy \citep{2013MNRAS.429.2562T}. 
IHV provides an opportunity to explore both the physical properties of the AGN including the brightness temperature, and the possible origin of discrete plasma in the solar neighbourhood. However, such extreme variability is rare. 

\citet{1997ApJ...490L...9K} found the first extreme IHV, PKS 0405--385, in the southern IDV survey, among monitoring of 125 selected bright, flat-spectrum AGN with the Australia Telescope Compact Array (ATCA). 
Other extreme IHVs such as J1819$+$3845~\citep{2000ApJ...529L..65D}, PKS 1257$-$326~\citep{2003ApJ...585..653B}, and recently J1402$+$5347~\citep{2020A&A...641L...4O} were discovered serendipitously. 
Several IDV/IHV surveys have been conducted \citep{1984A&A...134..390R, 2001MNRAS.325.1411K, 2019MNRAS.489.5365K} including the large-scale microarcsecond scintillation-induced variability survey \citep[MASIV;][]{2003AJ....126.1699L}, but no other extreme variables have been identified. 

The Australian Square Kilometre Array Pathfinder \citep[ASKAP;][Hotan A.W. et al (in press), PASA]{2014PASA...31...41H, 2016PASA...33...42M} is a survey telescope equipped with phased array feeds (PAFs) on $36\times12$-m dishes providing a $\sim30$ deg$^2$ field-of-view (FoV), giving us a good opportunity to investigate the radio dynamic sky \citep{2013PASA...30....6M}. 
The good sensitivity and instantaneous (\textit{u,\,v}) coverage \newadded{(baseline ranges from 22m to 6440m)} allow us to explore model-subtracted images on short time-scales (e.g. 15-minute) over a typical 10-h observation, making it possible to search for rapid variables $\lesssim$ hours in the image plane. 
Compared with existing IDV surveys, which have been limited to monitoring hundreds of relatively bright, flat-spectrum AGN, a search with ASKAP can monitor tens of thousands of sources simultaneously, becoming an unbiased search for fast scintillators. 
Apart from scintillating sources, other rapid variables with timescales $\lesssim$ hours, e.g. radio flaring stars \citep{2019MNRAS.488..559Z} or pulsars \citep{2019ApJ...884...96K}, can also be detected in such a survey. 

In this paper we present a search for hour-timescale variables in observations conducted as follow-up of the LIGO gravitational wave event S190814bv \citep{2019ApJ...887L..13D}.
Section~\ref{sec:processing} describes the data processing strategies and quality assessment.
Section~\ref{sec:blind_search} provides the details of source finding, lightcurve extraction and characterization, and selected criteria for variability. 
In Section~\ref{sec:results}, we present the results of variables found by our method, including their multi-wavelength counterparts, lightcurves, and sky distribution. 
In Section~\ref{sec:discussion}, we analyse and discuss the properties of our sources and their associated screen, before drawing our conclusions in Section~\ref{sec:conclusions}. 

\section{Observations and Data Reduction}
\label{sec:processing}

\subsection{Observations}

A series of observations were conducted on a 30 deg$^2$ field centred on RA: 00$^h$50$^m$37.5$^s$, Dec: -25$^{\circ}$16$'$57.4$''$ (J2000) using 36 ASKAP antenna dishes, with the original purpose of searching for a radio counterpart of the gravitational wave event S190814bv \citep{2019ApJ...887L..13D}. 
The field was observed using 36 beams arranged in a closepack36 footprint\footnote{\newadded{See more details in ASKAP Science Observation Guide:  \url{https://confluence.csiro.au/display/askapsst/?preview=/733676544/887260100/ASKAP_sci_obs_guide.pdf}}} 
with beam spacing of 0.9$^{\circ}$ (Hotan A.W. et al (in press), PASA).
\newadded{Each ASKAP beam has a field of view $\sim 1.6^\circ$ full width at half maximum (FWHM) and is correlated independently, when imaged and combined in a mosaic the total field of view is $\sim30$ deg$^2$. }
The field was tracked for $\sim 10$ hours in each epoch at a central frequency of 945~MHz and a bandwidth of 288~MHz.
Seven epochs were observed on the dates given in Table~\ref{tab:obs_details}. The typical noise in each epoch is $\sim$ 35 $\mu$Jy/beam with a synthesized beam size of $\sim$ 12$''$ \newadded{(resulting from a maximum baseline of 6.4 km)}. 
Our search was conducted on epochs 1 to 4; the details of the additional three follow-up epochs are discussed in Section~\ref{sec:followup}.

\begin{table}
	\centering
	\caption{Details of ASKAP observations for each epoch, including epoch number, scheduling block ID (SBID), start time (UTC), and the duration of each observation. }
	\label{tab:obs_details}
	\begin{tabular}{cccc} 
		\hline
		Epoch & SBID & Start Time (UTC) & Duration (h) \\
		\hline
		1   & 9602  & 2019-Aug-16 14:11:22.9 & 10.5 \\
		2   & 9649  & 2019-Aug-23 13:43:54.6 & 10.5 \\
		3   & 9910  & 2019-Sep-16 12:09:33.2 & 10.5 \\
		4   & 10463 & 2019-Nov-07 08:45:10.2 & 10.5 \\
		\hline
		5  & 12704 & 2020-Apr-03 22:59:59.9 & 5    \\
		    &       & 2020-Apr-04 22:55:15.7 & 10.5 \\
		6  & 13570 & 2020-Apr-29 21:41:10.6 & 10   \\
		7  & 15191 & 2020-Jul-03 17:01:26.4 & 10   \\
		\hline
	\end{tabular}
\end{table}

\subsection{Data Reduction}
\label{sec:data_reduction}

We reduced the data using the \texttt{ASKAPsoft} pipeline \citep{2017ASPC..512..431W} as described in \citet{2019ApJ...887L..13D} except for the epoch 5 observation as it failed half-way through the first day and was restarted the following day. To ensure sufficient $(u,v)$ coverage, the data from both days were first combined and then reduced with the \texttt{ASKAPsoft} pipeline. 

\subsubsection{Model-subtracted images}

We reduced the data using the Common Astronomy Software Applications package \citep[CASA;][]{2007ASPC..376..127M}. 
We imaged and processed each beam separately. 
For each of the 36 beams, we made an independent reference model image using multi-scale multi-frequency synthesis with two Taylor terms \newadded{(to allow the spectral curvature of sources to be modelled in the deconvolution process)} on the visibilities of epoch 1 using the \texttt{tclean} task in CASA. 
We performed a deep clean (10\,000 iterations) using Briggs weighting with robustness of $-0.5$ \newadded{to provide a compromise between resolution and sensitivity through visibility weighting} (and scales of 0, 5, 15 and 25 pixels to account for extended sources), and achieved a residual RMS and a final residual peak of about 40~$\mu$Jy/beam and 300~$\mu$Jy/beam respectively. 
A cell size of 2.5 arcsec and a large image size of 10\,000 $\times$ 10\,000 pixel were chosen so as to include the bright, extended object NGC~253 in the image, reducing possible sidelobe effects. 
We excluded five beams which contained the bright galaxy NGC 253 in the primary beam, given the possible adverse effects on variability search. 

We then converted the reference model images to model visibilities for each of the epochs separately.  
After that, phase self-calibration was performed on each of epochs with solution interval of 1 minute. 
We noticed the bright sources in epoch 4 have a flux scale error of about 5\% higher than them in other epochs. 
Considering the epoch 4 was observed two months after the other three (and observed during summer), there might be thermal effect causing gains to change. 
We applied amplitude self-calibration on epoch 4 data, also with solution interval of 1 minute, to correct the flux scale (see Fig.~\ref{fig:flux_scaling}). 
After that, we subtracted the model visibilities from the calibrated visibilities of each epoch. 
Finally, we imaged the model-subtracted visibilities in 15-minute time-steps using the same weighting parameters as before, generating 43 model-subtracted images each epoch. 
\newadded{Each beam was imaged over 3\,000 $\times$ 3\,000 pixels (2.1 $\times$ 2.1 deg square), which is about 1.5 times the diameter of the primary beam. }
Since models have been subtracted from visibilities, we did not apply any deconvolution in the 15-minute images.

\subsubsection{Image quality}
\begin{figure}
	\includegraphics[width=\columnwidth]{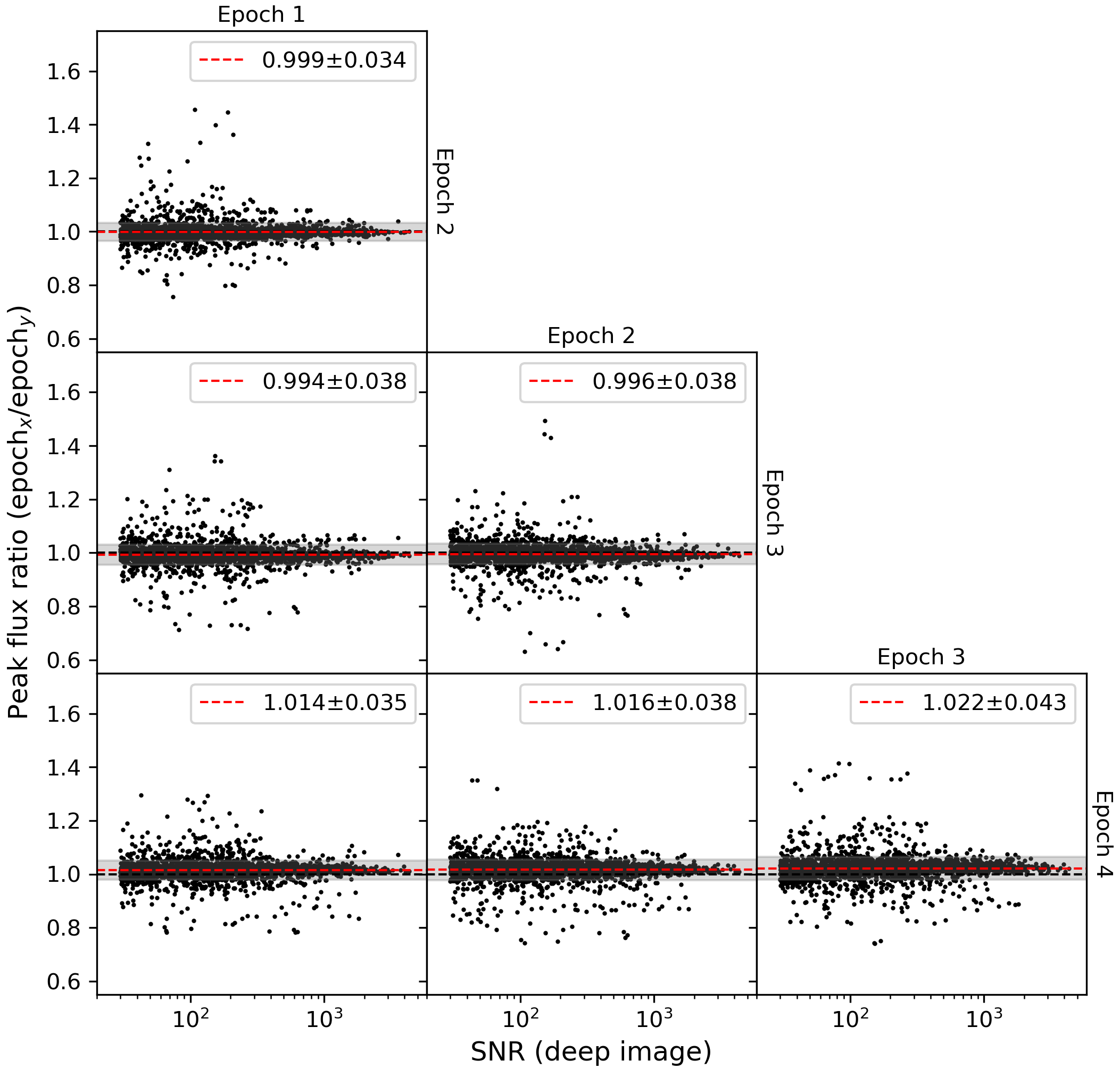}
    \caption{The peak flux density ratio of bright, compact sources for each epoch pair. The peak flux density of a selected source in each epoch was calculated by averaging data points in the light-curve of a given epoch.  The red dashed line shows the mean peak flux ratio of two epochs, and the grey shadow is the standard deviation. }
    \label{fig:flux_scaling}
\end{figure}

The overall astrometric accuracy and flux scale was evaluated by \citet{2019ApJ...887L..13D}. 
We evaluated the flux density stability of each epoch using a set of $\sim$ 3,000 bright compact sources selected on the following metrics:
\begin{enumerate}
    \item signal-to-noise ratio (SNR) $>30$;
    \item the ratio of integrated flux to peak flux $<1.2$;
    \item separation from beam centre $<0.45$ degree. 
\end{enumerate}
The average epoch to epoch flux density ratio was consistent to 1.0 with $\sim$ 4 percent uncertainty (see Fig.~\ref{fig:flux_scaling}). 

The typical rms noise in each model-subtracted image is $\sim$ 200$~\mu$Jy/beam. 
\newadded{We compared the rms noise of each 15-min image, and found} the noise varies throughout each observation as a result of elevation effects and variations in solar and radio-frequency interference. 
The relatively high rms at the middle and the end of observation is due to shorter integration times in those samples\footnote{Note that the middle observation is short as the roll axis of telescope had to unwrap. }.
The rms noise for different beams and epochs are generally consistent; beams with higher rms noise are located on the edge of the field or near NGC~253. 

\section{Search for Variability}
\label{sec:blind_search}

To identify highly variable sources on timescales of hours we conducted a search for variations within each epoch of our observations.

\subsection{Source detection and light-curve extraction}
\label{sec:source_dectection}

We generated a source catalogue from the 10.5-hour deep image of epoch 1 using \texttt{Aegean} \citep{2012MNRAS.422.1812H, 2018PASA...35...11H}. The built-in package \texttt{BANE} was used for estimating background and rms noise. 
Each beam was processed independently. We found about 1\,300 sources per beam (with image size of 2.1 $\times$ 2.1 square degrees) at 6$\sigma$ threshold, and most sources were detected at least twice on neighbouring beams (except for sources located on the edge of the processed field). We detected 40\,859 sources in total on all of processed beams. 

For each detection we extracted light-curves using the following steps:
\begin{enumerate}
    \item Obtained the peak flux density $S_\mathrm{deep}$ and fitted source position from the deep image catalogue;
    \item Measured peak flux density $S_{i, \mathrm{diff}}$ and rms noise $\sigma_i$ on $i$th model-subtracted image using the fitted position, resulting in 43 data points with errors per epoch;
    \item Added back the peak flux density from the catalogue for each data point, i.e. $S_i = S_{i, \mathrm{diff}} + S_\mathrm{deep}$, to get a light-curve. 
\end{enumerate}
This resulted in 163\,436 lightcurves (one per source per epoch) as inputs to our variability analysis.

\subsection{Variability measures}
\label{sec:variability_measures}

We used the modulation index to characterise the magnitude of variability, defined as
\begin{equation}
    m = \frac{\sigma_s}{\bar{S}}
    \label{eq:md}
\end{equation}
where $\bar{S}$ is the weighted mean flux density defined as
\begin{equation}
    \bar{S} = \frac{\sum_{i=1}^{n} \left(\frac{S_i}{\sigma_i^2}\right)}{\sum_{i=1}^{n} \left(\frac{1}{\sigma_i^2}\right)}
    \label{eq:weighted_mean_flux}
\end{equation}
and $\sigma_s$ represents the standard deviation of flux density of the light-curve. 
As mentioned in \citet{2014MNRAS.438..352B}, the modulation index is strongly dependant on the detection threshold of a source, so should be used in conjunction with the chi-squared value.

Following \citet{2011MNRAS.412..634B}, we used chi-square $\mathrm{\chi_{lc}^2}$ to measure the significance of random variability for light-curves. The calculation is based on the following expressions
\begin{equation}
    \mathrm{\chi^2_{lc}} = \displaystyle\sum_{i=1}^{n} \frac{\left(S_i - \bar{S}\right)^2}{\sigma_i^2}
    \label{eq:chisquare}
\end{equation}
where $\bar{S}$ represents the weighted mean flux density calculated by equation~(\ref{eq:weighted_mean_flux}), $S_i$ is the $i$th flux density in the light-curve obtained using above method (see description in Section~\ref{sec:source_dectection}), $\sigma_i$ is the estimated rms noise on $i$th measurement, $n$ is the total number of measurements in the light-curve ($n=43$ in this case for each detection on each epoch). We also calculated reduced chi-square $\mathrm{\chi_{red}^2}$ as
\begin{equation}
    \mathrm{\chi_{red}^2} = \frac{\mathrm{\chi^2_{lc}}}{n-1}
\end{equation}
Under the null hypothesis, the value of $\mathrm{\chi^2_{lc}}$ are expected to follow the distribution $\mathrm{\chi^2_{T}}$ with $n-1$ degrees of freedom. We calculated the probability of variability $P\left(\mathrm{\chi^2_{lc}}\right)$ for each light-curve using $\mathrm{\chi^2_{T}}$ cumulative distribution function (CDF). We considered light-curves with a probability above 3$\sigma$ significance level as a variable.

\subsection{Variability searches}
\label{sec:variability_searches}

\begin{figure}
	\includegraphics[width=\columnwidth]{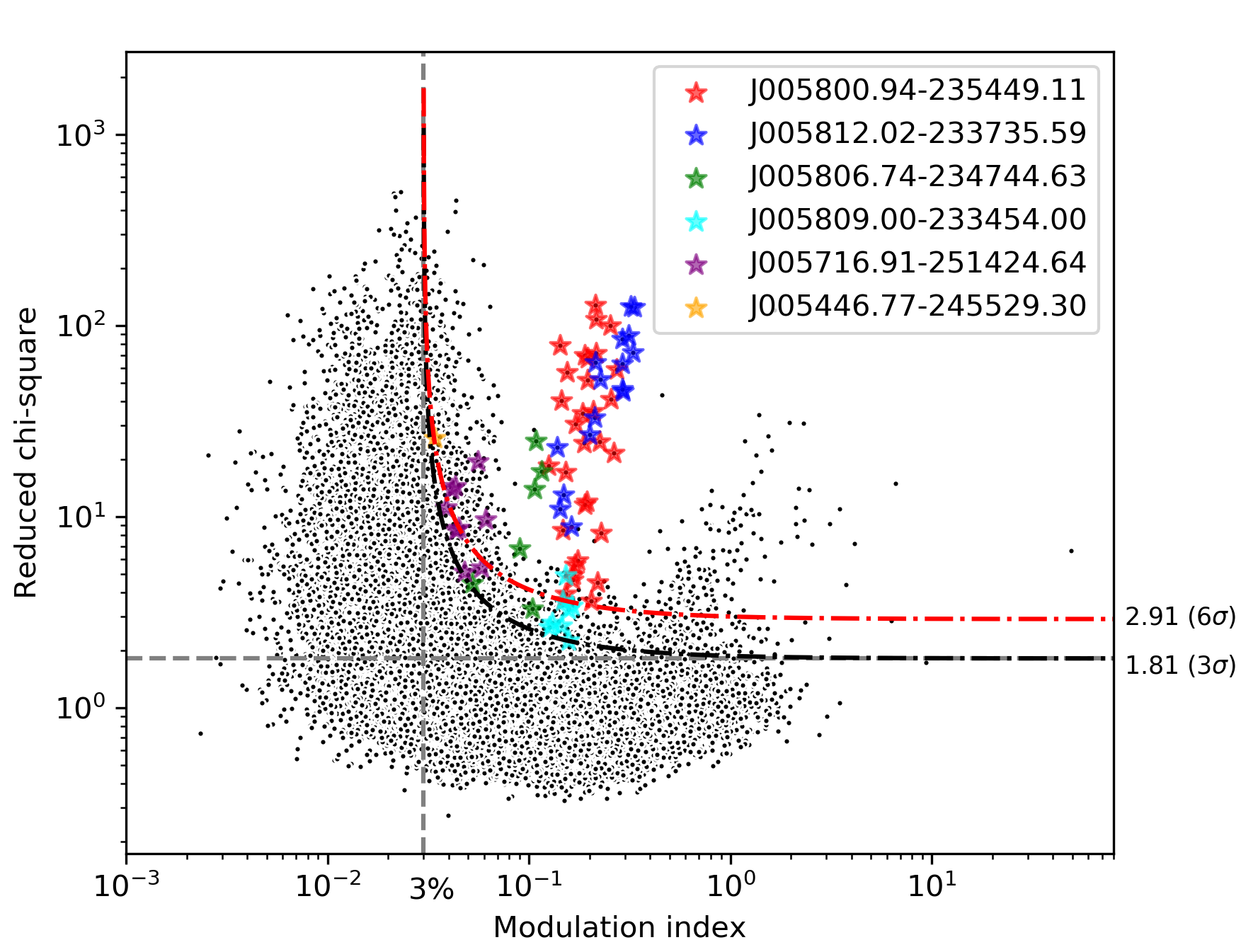}
    \caption{Distribution of modulation index and reduced chi-square for all 163\,436 light-curves. Each unique source corresponds to multiple light-curves detections from neighbouring beams and/or different epochs. We searched for sources with $\mathrm{\chi_{red}^2}$ and $m$ located on the top and right of the red dashed line (6$\sigma$) and the black dashed line (3$\sigma$) respectively, six highly variable sources were found (their chi-squared value all exceed the $6\sigma$ threshold in at least one epoch) in this field. The six variables, corresponding to multiple light-curves, were marked as stars with different colours. Another group of detections above the threshold (with modulation index about 1) are all false candidates near bright sources. }
    \label{fig:dis_chisq_md}
\end{figure}

As is customary, we selected highly variable source candidates as the outliers in the $m-\mathrm{\chi_{red}^2}$ plot shown in Fig.~\ref{fig:dis_chisq_md}. 
A stricter threshold was used for producing the list of candidates by finding all sources with 
\begin{enumerate}
    \item both chi-square and modulation index higher than the red dashed line on the $m-\mathrm{\chi_{red}^2}$ plot (see Fig.~\ref{fig:dis_chisq_md}), expressed as
    \begin{equation}
        \left(\frac{\mathrm{\chi_{red}^2}}{2.91}-1\right) \left(\frac{m}{3\%}-1\right)>1
        \label{eq:variability_metric}
    \end{equation}
    and we note that $\mathrm{\chi_{red}^2}=2.91$ corresponds to a $6\sigma$ significance level;
    \item separation from beam centre $<0.8$ degree.
\end{enumerate}

This resulted in 86 unique sources with 178 lightcurves (some sources were present in neighbouring beams and hence counted multiple times). 
All candidates were visually checked and some of them were rejected based on the following criteria: 
\begin{enumerate}
    \item Sources that were sidelobes of a bright source;
    \item Sources were extended, or had multiple components;
    \item Sources were coincident with imaging artefacts;
    \item Sources were not detected as variable in their main beam (i.e. the beam with smallest separation from beam centre to the source position). 
\end{enumerate}
Our final candidates have at least one detection at every epoch, and have to be detected at least once in their main beam. 
We found six highly variable sources that satisfied these criteria (marked in Fig.~\ref{fig:dis_chisq_md}).

We then searched again with less strict criteria by considering a $3\sigma$ chi-square threshold (i.e. the black dashed line in Fig.~\ref{fig:dis_chisq_md}), resulting total of 976 light-curves. However visual inspection ruled out all of the new candidates.

\newadded{As indicated before, the modulation index is highly dependant on the SNR of the source. We note that the large population with high $\mathrm{\chi_{red}^2}$ but low $m$ are bright sources $\geq 30$ mJy, and the population with high $m$ and low $\mathrm{\chi_{red}^2}$ are faint sources $\leq 0.5$ mJy (a majority of which are false detections near bright sources). For the well-behaved variables we identified, the modulation index and reduced chi-square of the source should be correlated following a power-law index of 2 (as roughly seen in Fig.~\ref{fig:dis_chisq_md}). This shows, from another aspect, that they are genuinely different and variable i.e. not due of some statistical coincidence. }

\section{Results}
\label{sec:results}

\begin{figure*}
	\includegraphics[width=2\columnwidth]{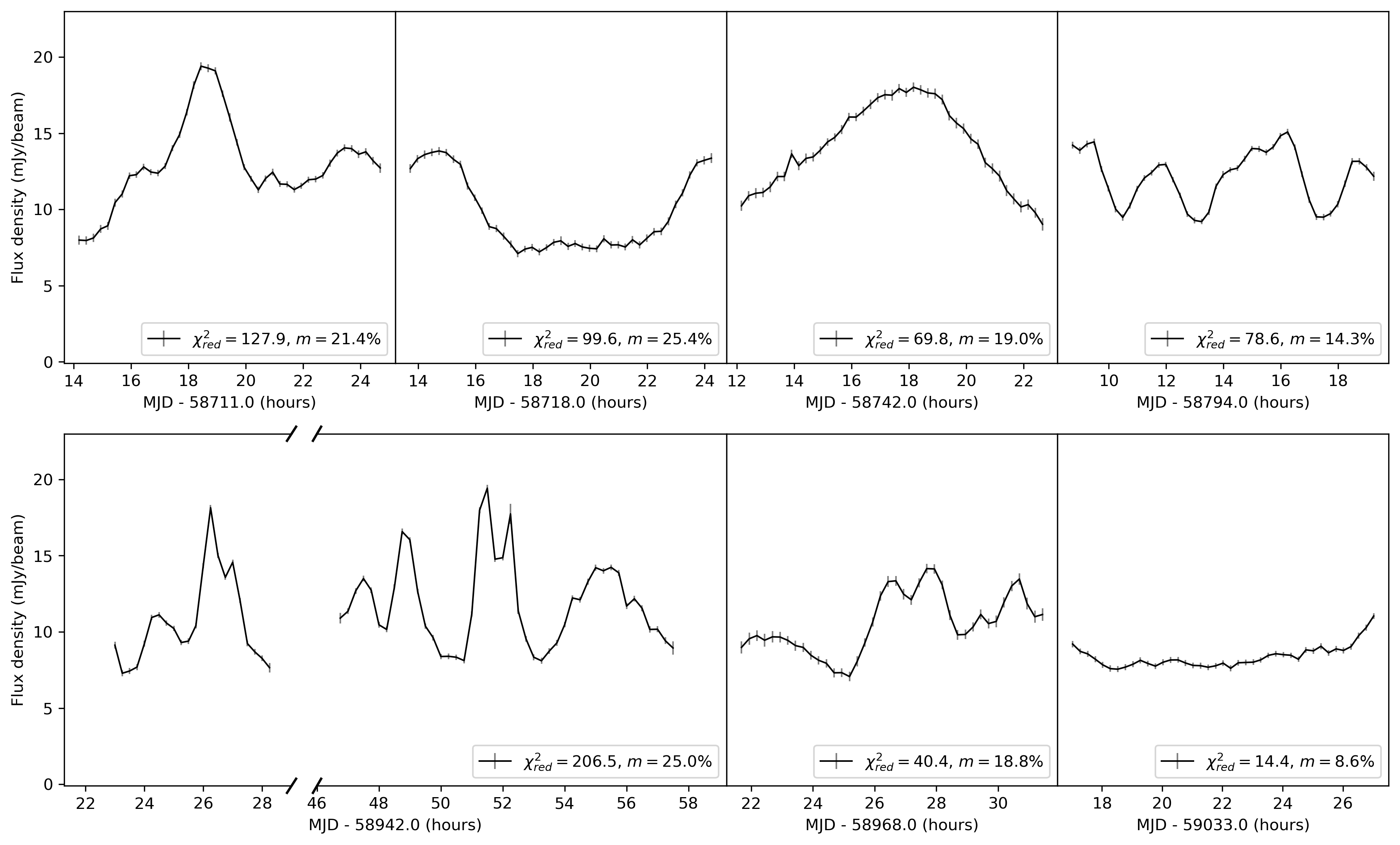}
    \caption{Light-curves in all epochs of source J005800.94$-$235449.11, one of the two extreme variables in our results. The light-curve in each epoch is measured in the main beam of the source after primary beam correction, \newadded{and the errorbar represents rms noise $\sigma_i$}. }
    \label{fig:lc_var_J005800}
\end{figure*}

\begin{figure*}
	\includegraphics[width=2\columnwidth]{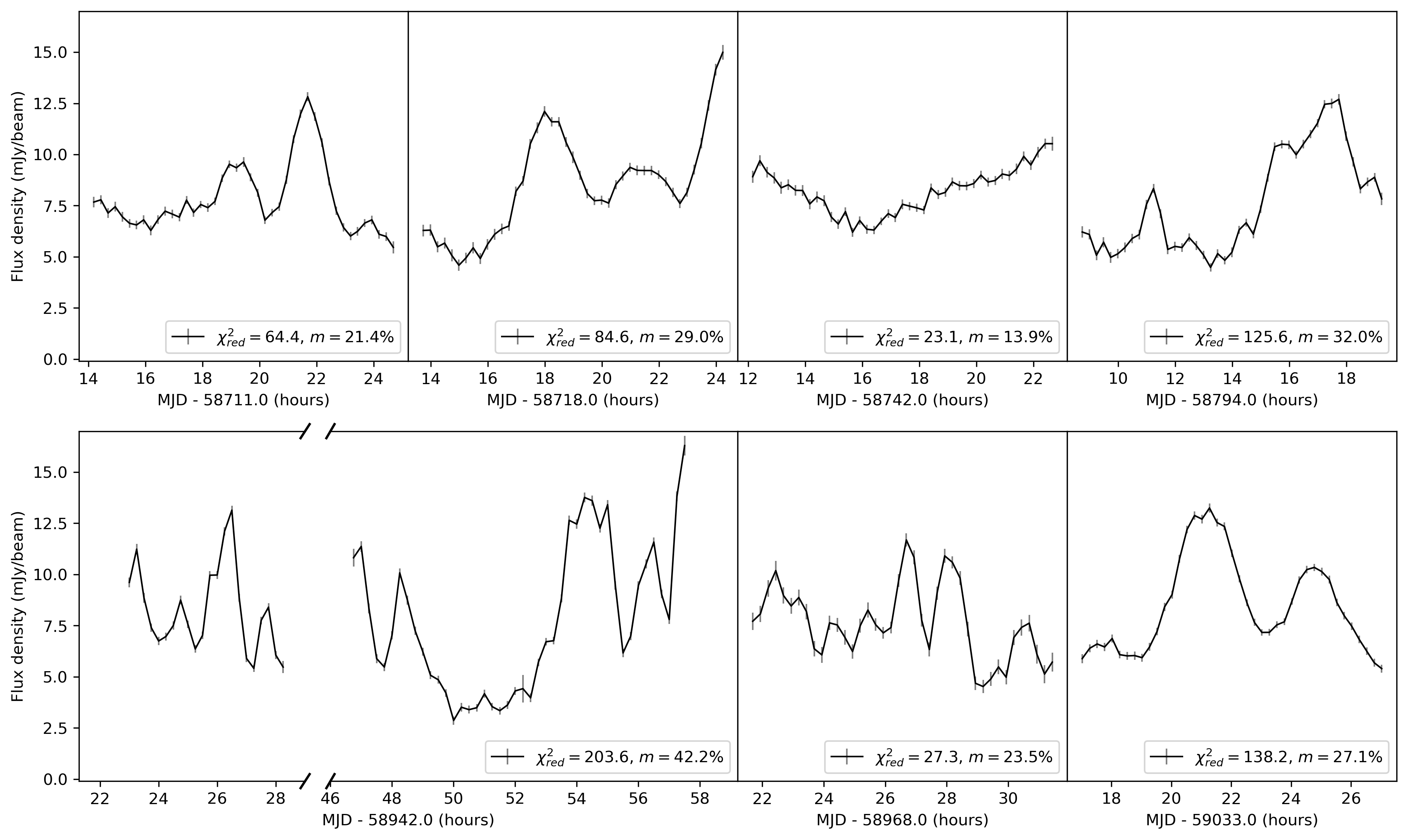}
    \caption{Light-curves in all epochs of source J005812.02$-$233735.39, another extreme variable in our lists. Details as in Fig.~\ref{fig:lc_var_J005800}. }
    \label{fig:lc_var_J005812}
\end{figure*}

We found six rapid variables, whose properties are listed in Table~\ref{tab:candidates}. 
Two of them, J005800.94$-$235449.11 and J005812.02$-$233735.39, are extreme variables, with reduced chi-square larger than 3.8 ($8\sigma$) in each epoch and a typical modulation index of $\sim 25\%$. 
Their variability timescales are as short as tens of minutes in some epochs. 
Their light-curves in all epochs (including three follow-up ones), are given in Fig.~\ref{fig:lc_var_J005800} and Fig.~\ref{fig:lc_var_J005812}. 
Light-curves of the four other variable sources, and movies of 15-min model-subtracted images can be found in Appendix~\ref{sec:details_variables}.

We searched for counterparts of these variables at other wavelengths.
For the radio band we used the NRAO VLA Sky Survey \citep[NVSS 1.4 GHz;][]{1998AJ....115.1693C}, Rapid ASKAP Continuum Survey \citep[RACS;][]{2020PASA...in..press} at 888 MHz, and the Very Large Array Sky Survey \citep[VLASS 3 GHz;][]{2020PASP..132c5001L}. 
We measured their flux density in RACS using \texttt{Selavy} \citep{2012PASA...29..371W}, and their peak flux density in VLASS quick look images~\citep{2020RNAAS...4..175G}. 
We also checked the Australia Telescope 20 GHz Survey \citep[AT20G;][]{2010MNRAS.402.2403M} as the most compact sources might appear there (e.g. sources with inverted spectra as the sensitivity of AT20G is about 40 mJy), but found no counterparts within radius of 10 arcsec. 
We used Vizier to search for Wide-ﬁeld Infrared Survey Explorer \citep[WISE;][]{2010AJ....140.1868W} and Dark Energy Survey~\citep[DES;][]{2018ApJS..239...18A} counterparts within 5 arcsec. 
Five sources had WISE counterparts, and their infrared colours suggest that they are AGNs. 
Source J005446.77-245529.3 also has corresponded SIMBAD ID of 2FGL~J0055.0-2454, and is identified as a BL Lac object \citep{2015Ap&SS.357...75M}. 
The details about multi-wavelength counterparts are listed in Table~\ref{tab:candidates}. 

\subsection{Sky distribution of variable sources}

\begin{figure*}
	\includegraphics[width=2\columnwidth]{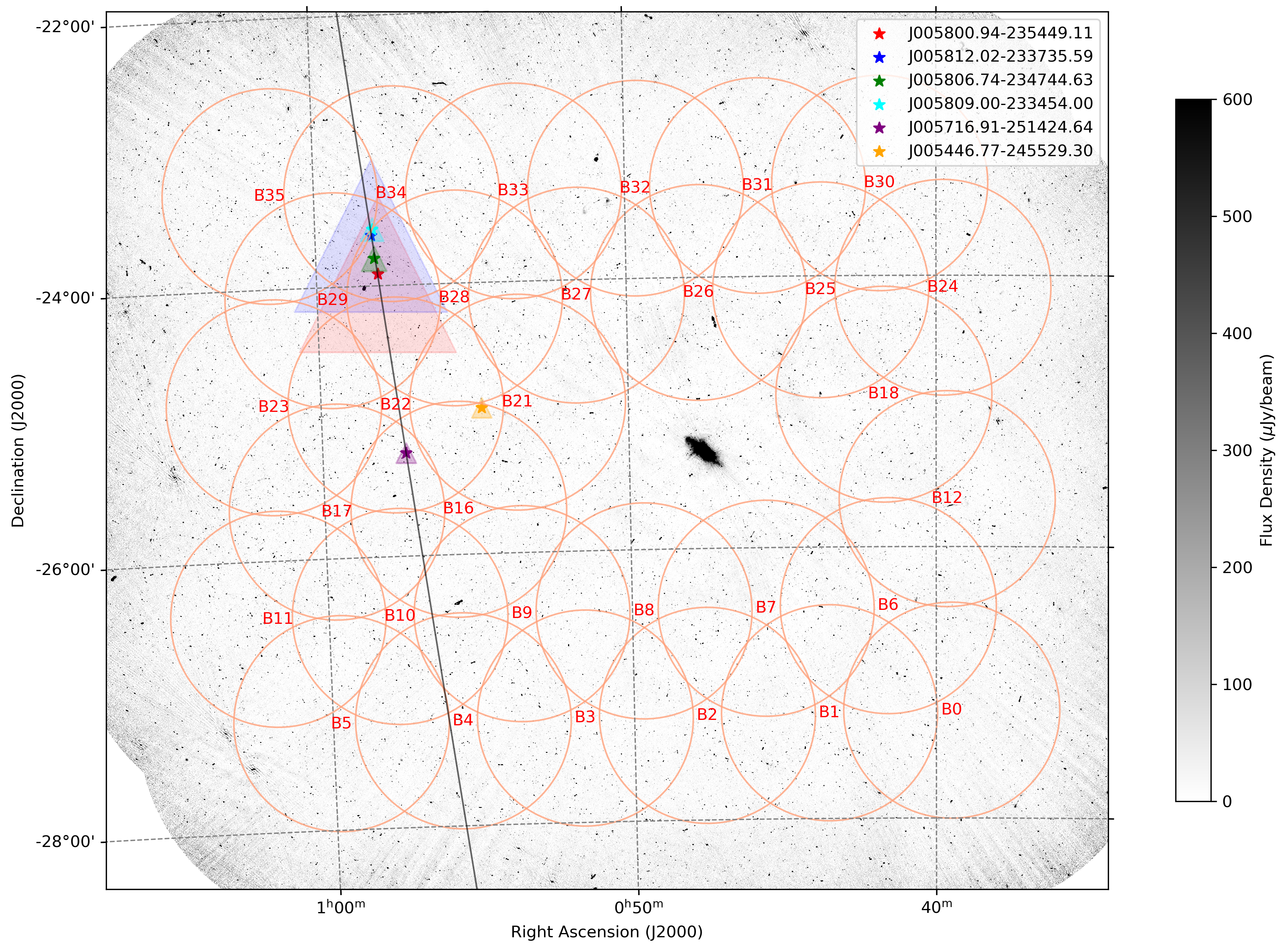}
    \caption{Sky distribution of variables and processed beams in the mosaiced deep image \newadded{(made by combining separate images of all beams)} of epoch 1 (with a zoom in version in Fig.~\ref{fig:dis_variables_zoomin}). We processed 31 of 36 beams (marked as orange circles with beam numbers in the centre, \newadded{and the diameter of each circle is consistent with the FWHM of the primary beam}), excluding 5 beams containing the bright object NGC 253 (i.e. beam 13, 14, 15, 19, and 20). The six highly variable sources we detected are marked as stars, and the size of the triangle shadow is proportional to the variability metrics, calculated by $\mathrm{\chi_{red}^2}\times m$. Source J005800.94-235449.11 (red) and source J005812.02-233735.39 (blue) are two extreme variable sources in our results. The black dashed line represents the best fitted great circle on the sky. \newadded{We note that the field is toward South Galactic Pole. }}
    \label{fig:dis_variables}
\end{figure*}

\begin{landscape}
\begin{figure}
	\includegraphics[width=\columnwidth]{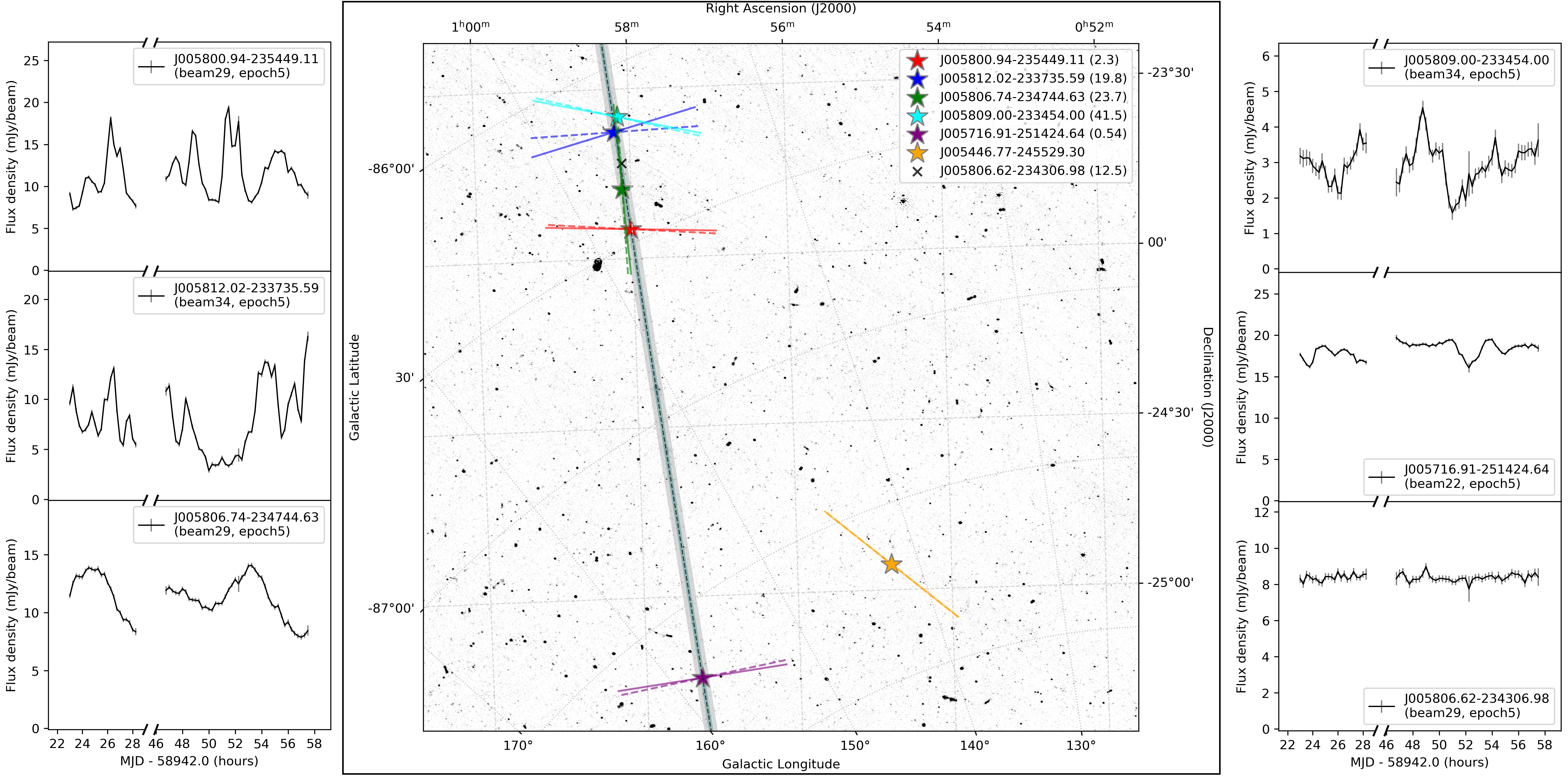}
    \caption{Magnified sky distribution of variables with anisotropy orientation. As in Fig.~\ref{fig:dis_variables}, the black dashed line is the best fitted great circle of the five variables, with offsets listed in the legend (unit of arcsec). The green shadow and grey shadow represent $1\sigma$ and $3\sigma$ area respectively. The light-curves of the five variables in epoch 5 were included. We also marked a non-variable source ("x" marker) with $<1\sigma$ separation to the line. The solid and dashed line in the sources are their orientations of the anisotropy, with 2D general model and 1D infinite model respectively (see details in  Section~\ref{subsection:kinematicanalysis}). }
    \label{fig:dis_variables_zoomin}
\end{figure}
\end{landscape}

An unexpected result of our analysis is the sky distribution of these highly variable sources. 
As shown in Fig.~\ref{fig:dis_variables}, five sources (except for the brightest J005446.77$-$245529.30) are in a linear arrangement on the sky, spanning approximately 1.7 degrees. 
To constrain how tight their distribution is to a line, we used least squares fitting of a great circle through the positions of five variables on the sky \citep{10.1145/367436.367478,10.1145/366678.366692}. 
To simplify the calculation, we used a unit normal vector to represent the plane of the best fitted great circle, and obtained the coordinates of $\alpha=287.422^{\circ}$, $\delta=6.547^{\circ}$ (J2000). 
The standard deviation of the source positions from the fitted projection line is $\sim23''$ (see Fig.~\ref{fig:dis_variables_zoomin}). 
We calculated the probability of this being a chance alignment by running 1000 trials of randomly choosing five compact sources in the field, and found the possibility of alignment with standard deviation $<1'$ less than $0.1\%$. 

We then investigated whether any instrumental or observational issues could be causing this effect. However, our analysis showed this was unlikely for the following reasons:

\begin{enumerate}
    \item The variables are compact in nature with $S_\mathrm{int}/F_\mathrm{peak} < 1.2$. Hence the variations do not seem to be caused by image artefacts or side-lobes of bright sources; 
    \item Variable behaviour was detected in every beam containing each source (the main beam and neighbouring beams); 
    \item The sources exhibit different variability behaviours to each other, and in each different epoch; 
    \item The sources are not particularly bright, therefore it is unlikely caused by calibration problems which should affect brighter sources more; 
    \item They are the only rapid variables along the line. Fig.~\ref{fig:dis_variables_zoomin} shows a nearby non-variable source J005806.62-234306.98, with similar flux density. It is difficult to explain how a systematic problem could affect only a few sources and none of the surrounding ones.
\end{enumerate}

The remaining possibility which might cause irregular variations for specific sources is some kind of unusual optics effect, given that the primary beam (and the footprint) didn't rotate and shift in those epochs, e.g. sources in specific positions might be wobbling in and out of peculiar shadows. 
We conducted a follow-up observation (epoch 6) with a rotated footprint and shifted phase centre to review this possibility, as described in the next section. 

\subsection{Follow-up observations with ASKAP}
\label{sec:followup}

Our original search for rapid variables was conducted on the first four epochs of data (i.e. SBID 9602, 9649, 9910 and 10463). 
We then observed three follow-up epochs for this field: the observation details are given in Table~\ref{tab:obs_details}. 

Epochs 5 and 7 used the same observing parameters as previous four epochs, with a field centred on $\alpha=00^h50^m37.5^s$, $\delta=-25^{\circ}16'57.4''$ using 36 beams in a closepack36 footprint at the central frequency of 945 MHz. 
In epoch 5 the field was tracked for 15.5-hour since the first observation failed after 5 hours and was restarted the next day. 
We obtained light-curves for the variable sources using the same method described in Section~\ref{sec:processing} and \ref{sec:blind_search}, using the model made from the Epoch 1 visibilities. 

Epoch 6 used different observing specifications to identify whether an instrumental or analysis effect was causing the observed linear arrangement on the sky.
The same observing frequency and footprint arrangement was used, but the phase centre was shifted to $\alpha=00^h58^m00^s$, $\delta=-23^{\circ}45'00''$ (nearer the cluster of variables). 
We also rotated the footprint by 67.5 degrees (an arbitrary amount), to give a different beam distribution on the sky. 
This epoch was tracked for 10-hour and the data reduction followed the same steps in Section~\ref{sec:data_reduction}. 
We used and subtracted the model made from the Epoch 6 visibilities itself since the field was shifted. 
We then searched for candidates using the same approach described in Section~\ref{sec:variability_searches}. 
We detected all of those rapid variables again except for source J005716.91$-$251424.64, which was brighter in this epoch and established weaker variability with $m<3\%$. 
No other rapid variables were detected in this region. 

The results from Epoch 6 (i.e. the detection of our variables again and non-detection of others) ruled out the possibility of technical issues causing the observed phenomenon. 
We therefore concluded that these variables, as well their unusual sky distribution, are of astrophysical origin. 

\begin{landscape}
\begin{table}
	\centering
	\caption{Properties of variable sources, including RA ($\mathrm{\alpha_{J2000}}$), DEC ($\mathrm{\delta_{J2000}}$) in J2000, galactic longitude ($l$) and latitude ($b$) in degree, peak flux density measured on the deep image of epoch 1, modulation index ($m$), \newadded{reduced chi-squared value ($\mathrm{\chi_{red}^2}$), scintillation rate ($R$; see details in Section~\ref{subsection:kinematicanalysis}), } number of detected beams (main beam in bold type), offsets from the best fitted great circle, and multi-wavelength counterparts if available.}
    \begin{flushleft}
        Note: The VLASS quick look images might have poor flux density accuracy according to their website. Source J005806.74--234744.63 was unresolved from a nearby brighter source in the original NVSS catalogue, but \citet{2001A&A...379...21Z} classified them as double radio sources and gave a separate flux measurement. In addition, source J005806.74--234744.63 has no match in the original WISE catalogue, and the WISE counterpart listed in the table is from unWISE \citep{2019ApJS..240...30S}, a catalogue based on WISE survey but with improved resolution and sensitivity. Source J005809.00-233454.00 was not identified in the original RACS catalogue (Hale et al., in preparation) but has a $4\sigma$ peak in RACS images.
    \end{flushleft}
	
	\label{tab:candidates}
	\begin{tabular}{llrrrrrr}
		\hline
		Name  & & J005800.94--235449.11 & J005812.02--233735.59 & J005806.74--234744.63 & J005809.00--233454.00 & J005716.91--251424.64 & J005446.77--245529.30 \\
		\hline
		$\mathrm{\alpha_{J2000}}$ & (deg) & 14.503897 & 14.550135 & 14.528069 & 14.537379 & 14.320473 & 13.694876 \\
		$\mathrm{\delta_{J2000}}$ & (deg) & $-23.913642$ & $-23.626548$ & $-23.79573$ & $-23.581653$ & $-25.240178$ & $-24.924807$ \\
		$l$ & (deg) & 148.07028097 & 146.86580877 & 147.62138918 & 146.44559058 & 158.03230313 & 141.92958546 \\
		$b$ & (deg) & $-86.45951442$ & $-86.17974985$ & $-86.34290489$ & $-86.14298906$ & $-87.70142128$ & $-87.67219925$ \\
		Peak flux & (mJy~beam$^{-1}$) & $11.53\pm0.040$	& $6.77\pm0.025$ & $9.71\pm0.016$ & $1.96\pm0.009$ & $15.86\pm0.026$ & $24.19\pm0.037$ \\
		$m$ (\%) & epoch 1 & 21 & 21 & 4.2  & 15 & 4.3 & 2.2 \\
		 & epoch 2 & 25 & 29 & 11 & 16 & 2.7 & 3.4 \\
		 & epoch 3 & 19 & 14 & 5.1  & 10  & 2.7 & 2.4 \\
		 & epoch 4 & 14 & 32 & 3.0  & 13 & 4.1 & 1.3 \\
		 & epoch 5 & 25 & 42 & 15 & 19 & 4.9 & 2.9 \\
		 & epoch 6 & 19 & 23 & 11 & 23 & 2.8 & 4.2 \\
		 & epoch 7 & 8.6  & 27 & 7.0  & 7.3  & 2.3 & 4.0 \\
		$\mathrm{\chi_{red}^2}$ & epoch 1 & 128 ($>8\sigma$) & 64 ($>8\sigma$) & 3.8 ($7.9\sigma$) & 4.9 ($>8\sigma$) & 14 ($>8\sigma$) & 13 ($>8\sigma$)\\
		 & epoch 2 & 100 ($>8\sigma$) & 85 ($>8\sigma$) & 25 ($>8\sigma$) & 3.3 ($6.8\sigma$) & 4.5 ($>8\sigma$) & 26 ($>8\sigma$) \\
		 & epoch 3 & 70 ($>8\sigma$) & 23 ($>8\sigma$) & 1.9 ($3.2\sigma$) & 1.5 ($2.2\sigma$) & 3.4 ($7.2\sigma$) & 10 ($>8\sigma$) \\
		 & epoch 4 & 79 ($>8\sigma$) & 126 ($>8\sigma$) & 2.9 ($5.9\sigma$) & 2.7 ($5.4\sigma$) & 14 ($>8\sigma$) & 3.6 ($7.4\sigma$) \\
		 & epoch 5 & 207 ($>8\sigma$) & 204 ($>8\sigma$) & 73 ($>8\sigma$) & 8.7 ($>8\sigma$) & 23 ($>8\sigma$) & 20 ($>8\sigma$) \\
		 & epoch 6 & 40 ($>8\sigma$) & 27 ($>8\sigma$) & 11 ($>8\sigma$) & 3.7 ($7.5\sigma$) & 7.9 ($>8\sigma$) & 48 ($>8\sigma$) \\
		 & epoch 7 & 14 ($>8\sigma$) & 138 ($>8\sigma$) & 15 ($>8\sigma$) & 1.6 ($2.4\sigma$) & 5.9 ($>8\sigma$) & 50 ($>8\sigma$)\\
		$R\,(\mathrm{d}^{-1})$ & epoch 1& $22\pm 1$& $16\pm 2$& $20_{- 3}^{+ 4}$& $18\pm 1$& $14.3_{-0.7}^{+0.8}$& $20\pm 2$ \\
         & epoch 2& $15.1\pm0.9$& $13\pm 2$& $6.8_{-0.8}^{+0.9}$& $20\pm 2$& $17_{- 1}^{+ 2}$& $13.5\pm0.9$ \\
         & epoch 3& $9.9\pm0.8$& $2.4\pm0.5$& $13\pm 4$& $13_{- 4}^{+ 5}$& $12_{- 1}^{+ 2}$& $5.7_{-0.9}^{+1.0}$ \\
         & epoch 4& $34\pm 3$& $15\pm 2$& $11_{- 2}^{+ 3}$& $40\pm 4$& $14\pm 1$& $ 7\pm 2$ \\
         & epoch 5& $60\pm 2$& $49\pm 4$& $20.3\pm0.7$& $31\pm 2$& $45\pm 2$& $24\pm 2$ \\
         & epoch 6& $28\pm 3$& $43\pm 8$& $19\pm 2$& $17_{- 3}^{+ 4}$& $28_{- 3}^{+ 4}$& $39\pm 3$ \\
         & epoch 7& $ 8_{- 2}^{+ 3}$& $12\pm 3$& $ 6\pm 2$& $10_{- 6}^{+ 9}$& $17_{- 3}^{+ 4}$& $26_{- 3}^{+ 4}$ \\
		Detected beams & & 21, 22, 23, 28, \textbf{29}, & 28, 29, \textbf{34}, 35 & 28, \textbf{29}, 33, 34, 35 & 28, 29, \textbf{34}, & 16, 17, 21, \textbf{22}, 23 & \textbf{21} \\
		(main beam in bold type) & & 33, 34, 35 & \\
		Offset to the line & (arcsec) & 2.3 & 20 & 24 & 41 & 0.54 & - \\
		\hline
		VLASS (3 GHz)* & (mJy~beam$^{-1}$) & $3.96\pm0.12$ & $4.07\pm0.12$ & $3.38\pm0.12$ & - & $25.84\pm0.15$ & $17.76\pm0.15$ \\
		NVSS (1.4 GHz) & (mJy) & $8.7\pm0.5$ & 5.1 ± 0.5 & 5.0 ± 0.5* & - & 14.6 ± 1.1 & 24.1 ± 0.9 \\
		RACS (888 MHz) & (mJy) & $8.11\pm0.05$ & $8.06\pm0.06$ & $8.91\pm0.11$ & 1.4 mJy~beam$^{-1}$ & $15.54\pm0.21$ & $23.39\pm0.07$ \\ 
		 & & & & & ($4\sigma$ peak) & & \\
		WISE & & J005800.99--235448.0 & J005812.03--233735.6 & J005806.76--234744.86* & - & J005716.92--251424.4 & J005446.75--245529.1 \\
		DES & & - & J005812.02--233735.4 & J005806.74--234744.5 & - & J005716.87--251424.4 & J005446.74--245529.0 \\
		\hline
	\end{tabular}
\end{table}
\end{landscape}


\section{Discussion}
\label{sec:discussion}

\subsection{A filamentary screen}
\label{subsec:screen}

We found six rapid variables in a 30 deg$^2$ field, and five of them are in a linear arrangement on the sky within an angular width of 1~arcmin, spanning about 1.7 degrees. 
After ruling out the possibility of instrumental issues, the implication is that there is an astrophysical connection between the five aligned scintillators. 
A straightforward explanation is that their variation behaviours have the same origin, i.e. scintillation caused by foreground plasma in the form of a thin filament, several degrees long. 

To date, no direct multi-wavelength observations of the scattering medium responsible for extreme scintillation have been identified, meaning the geometric properties of such screens remain unknown. 
\citet{2014MNRAS.442.3338P} proposed thin, corrugated, reconnection sheets as scattering objects in the ISM, while \citet{2017ApJ...849L...3V} suggested the turbulent edge of an elliptical plasma globule as the cause of IHV J1819$+$3845.

Our results allow us to constrain the size of the screen by the sky distribution of non-scintillating sources in the field. 
We selected a group of 525 sources using the following criteria: 
\begin{enumerate}
    \item compact in nature, with $\mathrm{F_{int} \leq 1.2~F_{peak}}$; 
    \item spectral index $\alpha > -0.5$ (obtained from ASKAP in-band data) since flat-spectrum sources are typically more compact than steep-spectrum sources and therefore more likely to exhibit scintillation \citep{2008ApJ...689..108L}; 
    \item flux density $\mathrm{F_{int} > 2}$ mJy, ensuring enough signal-to-noise ratio for detection of rapid variability if it exists. 
\end{enumerate}
These non-scintillating sources indicate the absence of the scattering medium in those lines-of-sight. 
We therefore constrained the width of the filament to be between 1~arcmin (from rapid variables) and 4~arcmin (from non-variables in the selected sources). 
Fig.~\ref{fig:dis_diffuse} shows the sky distribution of non-scintillating sources we used to constrained the width. 
For the length of the screen, only a lower limit of 1.7 degree could be set, because there are no radio sources in our field within $\pm3\sigma$ of the best-fit line and lying to the North of the group of scintillators. 
We didn't find any flat-spectrum, compact sources located in the line between the five variables, so we have no evidence for patchiness in the distribution of the scattering plasma within the filament. 

This is the first time that multiple scintillators have been detected behind the same scattering screen.
Most previous surveys for IHV/IDV used bright targets distributed all over the sky \citep[e.g.][]{2003AJ....126.1699L} and so has limited utility in constraining the screen geometry. In deep searches of fields around known scintillators \citep[e.g.][]{2015A&A...574A.125D} radio telescopes with smaller fields of view would have difficulty recognising a structure as large as the one we have found. 

To estimate how reliably we can recognise similar, filamentary plasma screens we undertook the following test. We selected a thin, rectangular area placed at a randomly chosen location within our field, and with a randomly chosen orientation, and we counted the number, $n$, of compact sources (meeting the aforementioned criteria) lying within the chosen rectangle. This procedure was repeated 1\,000 times. To avoid possible edge effects in the statistics, the centres of these hypothetical screens were restricted to the central $4^{\circ} \times 4^{\circ}$ of the field. We define the discovery rate, $D(n)$, as the fraction of these hypothetical screens having $n$ such background sources. For a rectangle of length 1.7~degrees and a width of 1~arcmin we found: $D(0)=76.6$\%; $D(1)=20.1$\%; $D(2)=2.9$\%; and, $D(3)=0.4$\%. And increasing the width to 4~arcmin increased the discovery rates to: $D(0)=32.6$\%; $D(1)=34.8$\%; $D(2)=21.4$\%; $D(3)=7.8$\%; $D(4)=2.4$\%; and, $D(5)=0.8$\%. These rates are generally consistent with a Poisson distribution whose mean is the area of the screen multiplied by the source density ($\simeq 11\;{\rm deg^{-2}}$ over the central $30\;{\rm deg^2}$).

Even for the larger assumed width of 4~arcmin the rate $D(\ge 5)$ is very small, so the screen is likely to be significantly longer than the minimum possible value of $1.7^\circ$. In order to recognise a (thin, straight) filamentary geometry for a given scattering screen we would need to see at least three scintillators behind it, so the ratio $D(\ge 3)/D(\ge 1)$ is a gauge of how reliably we can do that. Based on the numbers given above (4 arcmin width), we expect to be able to recognise filamentary geometry in at least 16\% of cases (more if the screen is longer than $1.7^\circ$).

\subsection{Kinematic analysis}
\label{subsection:kinematicanalysis}
Variations in the flux due to interstellar scintillations arise as the telescope moves through the pattern of bright and dark patches projected by the plasma screen. The pattern drifts through the Solar system with a constant velocity -- which, in the case of extragalactic sources, is essentially the velocity of the screen transverse to the line of sight \citep{1998ApJ...507..846C}. As the velocity of the Earth changes through the year, the rate of scintillation changes. This annual modulation, or annual cycle, of the scintillation rate was instrumental in establishing the scintillation nature of the IHV phenomenon \citep{2000aprs.conf..147J, 2001ApJ...550L..11R} and can be used to determine the velocity of the screen along with the characteristic scale, degree of anisotropy and orientation of the scintillation pattern \citep[e.g.,][]{2001A&A...370L...9J,2003A&A...404..113D,2020A&A...641L...4O}.

We have determined the scintillation rate $R$ -- defined as the inverse of the flux autocorrelation function (ACF) half-width at half-maximum (HWHM) -- on all epochs by assuming the light curves to represent a Gaussian process and performing a global \tyomaadded{Markov Chain Monte Carlo (MCMC) fit of the parameters of the ACF, modelled as a damped cosine, for a reference epoch and time stretching factors for all other epochs}. The method, described in detail in \cite{2019MNRAS.487.4372B}, is designed to allow quantitative inference on the scintillation rate for epochs near standstills (when the Earth velocity is close to that of the screen) where traditional ACF HWHM estimates struggle due to very slow variations. It has been shown to produce results that closely follow the traditional ACF analysis on fast epochs, where comparison is possible \citep{2020A&A...641L...4O}. Unlike~\cite{2019MNRAS.487.4372B}, we used single light curves (per epoch, per source) due to observed relatively broadband nature of the scintillation relative to the ASKAP bandwidth.

\begin{figure*}
    \centering
    \includegraphics[width=58mm]{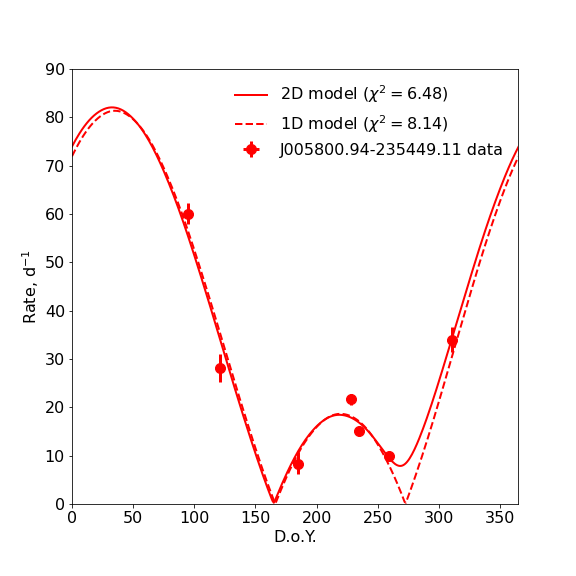}
    \includegraphics[width=58mm]{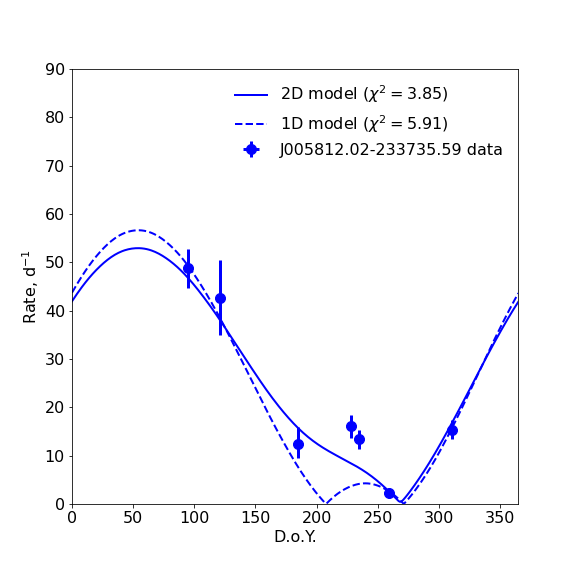}
    \includegraphics[width=58mm]{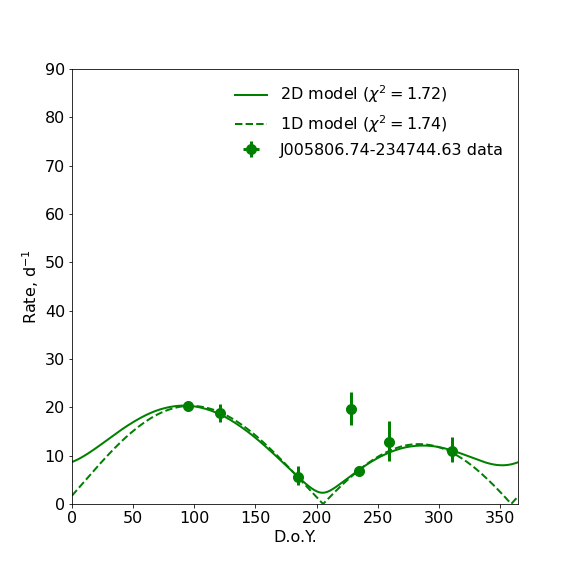}
    \includegraphics[width=58mm]{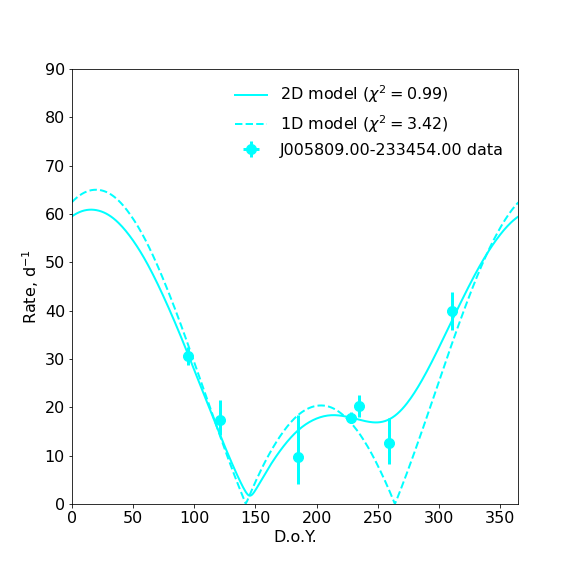}
    \includegraphics[width=58mm]{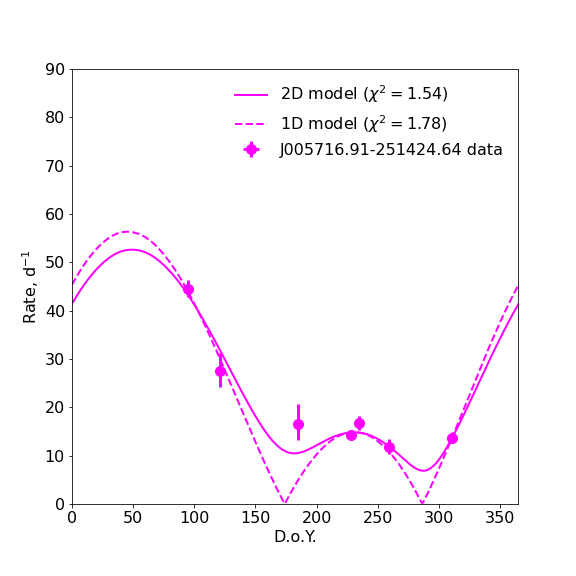}
    \includegraphics[width=58mm]{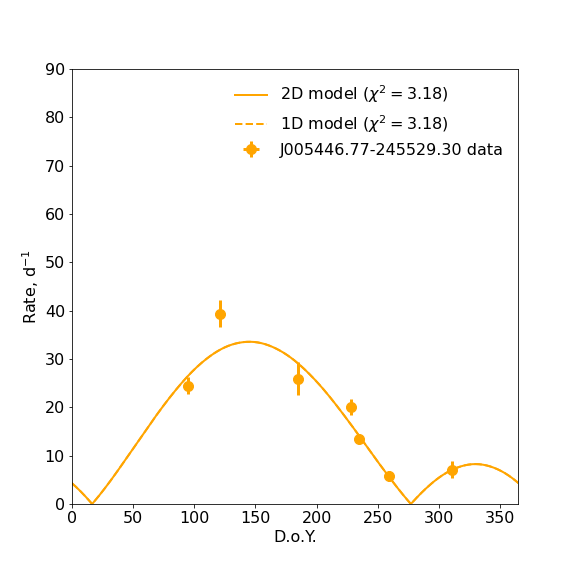}
    \caption{Variation of the scintillation rate through the year for six sources presented in this paper along with the best fit general (2D) and infinitely anisotropic (1D) models shown as solid and dashed line, respectively. The parameters of these models are given in Table~\ref{table:kinematicparameters}. Qualitatively, the measured cycles appear similar for the five sources observed along a single line while the one not on the line, J005446.77-245529.30, looks somewhat different. For J005446.77-245529.30, the 1D model provides the overall optimum.}
    \label{fig:rates}
\end{figure*}

Fig.~\ref{fig:rates} presents the rate estimates produced by our code along with the kinematic models that best fit those estimates. A general, finitely anisotropic model of the scintillation rate has five model parameters: two measurements of the spatial auto-covariance ellipse of the projected flux pattern (the light curve is assumed to represent a Gaussian process) along its principal axes $a_{\parallel,\perp}$, the orientation of the major axis of this pattern, $\mathrm{PA}$, and two components $v_\mathrm{screen}^{\parallel,\perp}$ of the projected screen velocity, as per 
\begin{eqnarray}\label{ratesquared}
R_i^2=\frac{\left(v^\parallel_{\oplus, i}-v^\parallel_\mathrm{screen}\right)^2}{a_\parallel^2}+\frac{\left(v^\perp_{\oplus,i}-v^\perp_\mathrm{screen}\right)^2}{a_\perp^2},
\end{eqnarray}
where ${\bf v}_{\oplus,i}$ is the Earth velocity on $i$-th epoch. We formed a $\chi^2$-like sum over the epochs using the mean values of the rate squared returned by the MCMC modelling as measurements and the difference between its $84$-th and $16$-th percentile as their uncertainties and performed a grid search for its minimum over $\mathrm{PA}$ and $v_\mathrm{screen}^{\parallel,\perp}$. The minimisation in $a_{\parallel, \perp}$ is a linear problem for this sum subject to $a^2$ positivity constraint. The derived parameters along with their uncertainties are summarised in Table~\ref{table:kinematicparameters}. \footnote{This treatment neglects correlations between the epoch estimates, which are not fully independent due to the global nature of the MCMC likelihood. Full account of this interdependence is very computationally expensive but we performed a smaller scale study taking it into account on a subset of MCMC output, which produced results very close to those presented here. \tyomaadded{We note that the uncertainties returned by our routine are likely underestimated due to the complex shape of the global likelihood, which might result in some corners of the parameter space to remain under-explored by the chain while artificially boosting the measure of those explored; this issue is discussed further in \citep{2019MNRAS.487.4372B}. The uncertainties of the kinematic model parameters are thus likely biased low.}}

\begin{table*}
    \centering
    \begin{tabular}{ccccccc}
        \hline
         Source& $\chi^2$ & $a_\perp,\,\mathrm{Mm}$ & $a_\parallel:a_\perp$ & $\mathrm{PA},^\circ$ & $v_\perp,\,\mathrm{km}\,\mathrm{s}^{-1}$ & $v_\parallel,\,\mathrm{km}\,\mathrm{s}^{-1}$ \\
         \hline\\
         J005800.94--235449.11 & 6.5 & $29\pm2$ &$14_{-5}^{+\infty}$ & $85\pm3$ & $-11\pm1$ & $-9\pm100$\\ & 8.1 & $30\pm2$ & & $87\pm2$ & $-10\pm1$ & \\ \\
         J005812.02--233735.59 & 3.8 & $59_{-3}^{+30}$ &$2.6_{-0.7}^{+40}$ & $92_{-17}^{+26}$ & $14_{-30}^{+13}$ & $26_{-10}^{+60}$\\ & 5.9 & $62_{-10}^{+70}$ & & $105_{-17}^{+45}$ & $18_{-30}^{+15}$ & \\ \\
         J005806.74--234744.63 & 1.7 & $160_{-80}^{+140}$ &$2.8_{-1.8}^{+\infty}$ & $2_{-2}^{+178}$ & $7\pm40$ & $30\pm100$\\ & 1.7 & $140_{-70}^{+160}$ & & $4_{-4}^{+176}$ & $7\pm40$ & \\ \\
         J005809.00--233454.00 & 1.0 & $33_{-5}^{+34}$ &$6.1_{-3.3}^{+24}$ & $75_{-7}^{+9}$ & $-9_{-10}^{+3}$ & $-12_{-100}^{+25}$\\ & 3.4 & $30_{-5}^{+200}$ & & $77_{-13}^{+15}$ & $-8_{-60}^{+15}$ & \\ \\
         J005716.91--251424.64 & 1.5 & $57_{-10}^{+50}$ &$13_{-11}^{+\infty}$ & $100\pm20$ & $13_{-22}^{+9}$ & $86\pm100$\\ & 1.8 & $49_{-3}^{+4}$ & & $97\pm5$ & $12\pm1$ & \\ \\
         J005446.77--245529.30 & 3.2 & $64_{-7}^{+17}$ &$>3.1$ & $50_{-5}^{+10}$ & $10_{-5}^{+4}$ & $0\pm100$\\ & 3.2 & $64_{-9}^{+24}$ & & $50_{-6}^{+5}$ & $10_{-3}^{+6}$ & \\
        \hline
    \end{tabular}
    \caption{General (finitely anisotropic, 2D) and extremely anisotropic (1D) model parameters that best fit the measured annual cycles. Optimisation was performed on a grid extending to $v_{\parallel,\perp}=\pm100\,\mathrm{km}\,\mathrm{s}^{-1}$with $1\,\mathrm{km}\,\mathrm{s}^{-1}$ step, $\pm100\,\mathrm{km}\,\mathrm{s}^{-1}$ in the uncertainty shows that the entire range is acceptable. Note that $\chi^2$ shown are the total, not reduced values; the 2D modelling has 2 effective degrees of freedom (7 epochs minus 5 model parameters) while for the 1D model this value is 4. For J005446.77-245529.30, the general optimum is achieved by the 1D model ($a_\parallel^{-2}\to0$). For J005806.74-2334744.63, $\mathrm{PA}$ is poorly constrained and the acceptable $v_\perp$ region is $\mathrm{PA}$-dependent. }
    \label{table:kinematicparameters}
\end{table*}

According to equation~(\ref{ratesquared}), amplitude of the $\parallel$ term is suppressed by the $1/a^2_\parallel$ factor and we therefore expect the $\perp$ components to be better constrained, which is confirmed in the table. In the infinite anisotropy limit, $a_\parallel:a_\perp\to\infty$, only the second term survives; this infinitely anisotropic model has only three parameters ($a_\perp$, $v_\mathrm{screen}^\perp$ and $\mathrm{PA}$) and was in fact shown to be sufficient when describing the annual cycles of other IHVs \citep{2003A&A...404..113D, 2003ApJ...585..653B, 2009MNRAS.397..447W, 2019MNRAS.487.4372B, 2020A&A...641L...4O}. Table~\ref{table:kinematicparameters} reveals that infinitely anisotropic model likewise provides an adequate description for most of the fast variables reported in the present paper and will in fact be preferred if the general model is penalised for its extra free parameters; where the general model is preferred its anisotropy degree is high. The orientation of the anisotropy major axis is displayed in Fig.~\ref{fig:dis_variables_zoomin}; Fig.~\ref{fig:localclouds} presents the two-dimensional view of the allowed screen velocities for all variables.

\subsection{Physical properties}
\label{subsec:physical_properties}

The scattering caused by the ionized ISM can be modelled with a power-law power spectrum of electron density fluctuations, associating spatial scale with phase structure --- it usually specified as isotropic Kolmogorov turbulence, whose power-law index is $\alpha=5/3$ \citep{1995ApJ...443..209A}.
In the case of point-like radio sources the character of the scintillations is then determined by the scattering strength, $U$, which depends on the observing wavelength $\lambda$ and the intrinsic amplitude of the turbulence \citep{2006ApJ...636..510G}.
To constrain our plasma filament, we first assumed a point-like model for all five sources in the line; in this limit their flux variations are expected to be similar. 
However, the observed modulation indices and variability timescales are very different from each other. 
One possible interpretation of this is that they are not point-like and have different source sizes $r_\mathrm{s}$. 
\citet{1992RSPTA.341..151N} indicated that a larger source size can lead to a reduction in the amplitude of variation and an increase in the scintillation timescale, which combination we have observed in our data. 
When introducing an extra variable $r_\mathrm{s}$, we need more constraints to fit the scintillation model. 
We therefore made dynamic spectra of these sources from ASKAP model-subtracted visibilities, with frequency resolution of 16 MHz and time resolution of 15 minute. 

The modulation indices of our two most extreme scintillators are close to unity, suggesting that they might be near the transitional scattering regime, i.e. scattering strength $U\sim 1$. 
We therefore used the fitting formulae given by \citet{2006ApJ...636..510G}, which are valid over a wide range of scattering strength including the transitional regime. (The description of \citet{2006ApJ...636..510G} is appropriate for isotropic Kolmogorov turbulence, whereas our kinematic analysis (Section~\ref{subsection:kinematicanalysis}) suggests significant anisotropy; there is, however, no comparable analysis for the anisotropic case.)
Their formulae describe a flux correlation, $W(U, r, r_\mathrm{s}, \eta)$, that is a function of the scattering strength $U$, spatial separation $r$ in the observer's plane, source size $r_\mathrm{s}$, and frequency difference $\eta$. 
In our data we find that the modulation indices are approximately constant across our observation band.
To identify appropriate models we constructed an ad hoc likelihood function for $W(U, r=0, r_\mathrm{s}, \eta=0)$, using modulation index epoch-to-epoch variation as an estimate of its uncertainty, and conducted a grid search to find the best $(\lambda_0, r_\mathrm{s})$ pairs, where $\lambda_0$ is the transition wavelength that is related to the scattering strength via $U = \left(\lambda/\lambda_0\right)^{(4+\alpha)/2}$. 

Our grid search revealed two branches of possible solutions that are broadly consistent with the behaviour of the modulation index. One is the weak-to-transitional regime where the transition frequency is within or just below the observing band. In this case the broad-band modulation is just below unity and the source size is only constrained to be less than a few Fresnel units. The other possible solution corresponds to strong scattering with a source size that is comparable to the diffractive scale. Diffractive scintillation is expected to be narrow-band, in the sense that the decorrelation bandwidth should be small compared to the observing frequency. However, our instrumental bandwidth is itself only $\sim30$\%\ of the observing frequency, and the observed in-band decorrelation factors ($0.1 - 0.6$) permit possible solutions with transition frequencies as high as $\sim10\,\mathrm{GHz}$.

With these considerations in mind the modelling results might be best summarised by stating that the transition frequency $f_0$ is likely to be close to $1\,\mathrm{GHz}$ but could be slightly lower or a few times higher. For Kolmogorov turbulence in the inertial range this transition frequency implies a scattering measure \citep[cf.~][]{1990ARA&A..28..561R,1991Natur.354..121C} of
\begin{eqnarray}
\mathrm{SM}=1.2\times10^{-4}\,\mathrm{m}^{-20/3}\,\mathrm{kpc}\,\left(\frac{f_0}{\mathrm{GHz}}\right)^{17/6}\left(\frac{D_\mathrm{screen}}{\mathrm{pc}}\right)^{-5/6}.
\end{eqnarray}
Assuming an outer scale that subtends an angle $\theta_\mathrm{out}$ on the sky, this scattering measure corresponds to a column density dispersion of
\begin{eqnarray}
\sigma^2_N=\left(5.4\times10^{15}\,\mathrm{cm}^{-2}\right)^2\,\left(\frac{f_0}{\mathrm{GHz}}\right)^{17/6}\left(\frac{D_\mathrm{screen}}{\mathrm{pc}}\right)^{5/6}\left(\frac{\theta_\mathrm{out}}{1\,\mathrm{'}}\right)^{5/3},
\end{eqnarray}
where we have used the apparent width of the filament as an estimate for the outer scale of the turbulence. If we also take the width of the filament as an estimate for the line-of-sight depth of the plasma then we obtain an estimate of the variance of the volume electron density fluctuations:
\begin{eqnarray}\label{eq:volumedensity}
\sigma^2_n=\left(1.7\,\mathrm{cm}^{-3}\right)^2\,\left(\frac{f_0}{\mathrm{GHz}}\right)^{17/6}\left(\frac{D_\mathrm{screen}}{\mathrm{pc}}\right)^{-7/6}\left(\frac{\theta_\mathrm{out}}{1\,\mathrm{'}}\right)^{-1/3},
\end{eqnarray}
and this result serves as an estimate of (the square of) the mean electron density.

Previous studies of scintillating sources have typically inferred a high source brightness temperature when variability is present on timescales of a few hours --- e.g. \citet{2002ApJ...581..103R} preferred $T_b\sim2\times10^{13}\;{\rm K}$ for the prototype IHV  PKS~0405$-$385. Such high brightness temperatures are thought to be rare in the radio source population, but appear preferentially amongst bright IHVs as a result of a strong selection bias: large amplitude scintillation requires a small angular size, so bright scintillators are likely to have high brightness temperatures. Our study, however, presents some of the faintest IHV sources that have ever been reported, two orders of magnitude fainter than most previous studies, so we expect our sources to exhibit much less bias. Moreover, the five scintillators located behind the filamentary screen are not accompanied by a larger number of non-scintillators in the same region. It therefore seems highly unlikely that those five IHVs all have high brightness temperatures. More likely they have brightness temperatures that are typical of AGN radio cores and thus lie in the range $10^{11}-10^{12}\;{\rm K}$ \citep{1998AJ....115.1295K}. \footnote{We also note that our scintillators don't seem to have the inverted radio spectra that are typical of the most compact synchrotron sources. Caution is necessary, however, as the fluxes shown in Table~\ref{tab:candidates} were measured at different epochs for different frequencies.}

\begin{figure}
\includegraphics[width=90mm]{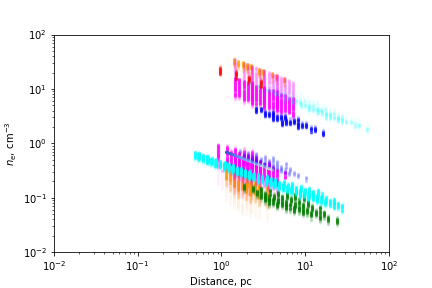}
    \caption{Acceptable models of the observed scintillation properties; sources are colour-coded as per Fig.~\ref{fig:rates}. Here our screen models are recast to electron density as a function of screen distance, making use of equation~(\ref{eq:volumedensity}). The two distinct groups of solutions, separated by a factor $\sim30$ in electron density, correspond to transitional scattering (lower branch) and strong, diffractive scattering (upper branch). We have restricted this plot to source brightness temperatures in the range $10^{11}-10^{12}\;{\rm K}$.}
    \label{fig:densitydistance}
\end{figure}

Each of our screen models implies -- through the ratio of source size to screen distance -- a particular source brightness temperature, so we have imposed $10^{11}\lesssim T_b({\rm K})\lesssim 10^{12}$ as an additional requirement on acceptable screen models. The resulting solution set is plotted in Fig.~\ref{fig:densitydistance} in the form of the implied electron density (from equation~(\ref{eq:volumedensity})) as a function of the distance to the screen. Because we have two distinct types of solution -- i.e. broad-band transitional scattering, and narrow-band diffractive scattering -- two separate branches are evident in Fig.~\ref{fig:densitydistance}, with most sources appearing in both branches. Our data do not allow us to decide between these two branches so we are only able to make rough estimates of the screen properties: $n_e\sim 1\;{\rm cm^{-3}}$, with a factor of 10 uncertainty; and, $D_\mathrm{screen}\sim 4\;{\rm pc}$, with a factor of 5 uncertainty. Although crude, these estimates suffice to illustrate the likely physical characteristics of the structure we are dealing with: it is a plasma filament with a length $\sim 10^{-1}\;{\rm pc}$, width $\sim 10^{-3}\;{\rm pc}$ and mass $\sim10^{-8}\;{\rm M_{\sun}}$ (assuming one proton per free electron).

We can also make an order-of-magnitude estimate of the number of such screens in the solar neighbourhood: assuming that this ASKAP field is representative, and allowing for a discovery efficiency of some tens of percent (see Section~\ref{subsec:screen}), we arrive at a volume density of $\sim 10\;{\rm pc^{-3}}$ --- a few times larger than that of ordinary stars.

Using the inferred physical properties given above we can estimate the geometric optical depth, $\tau$, of the screen population: considering only screens that are similarly local (i.e. within $\sim 4\;{\rm pc}$) we find $\tau\sim 4\times 10^{-3}$. We therefore anticipate that $\sim 0.4$\% of radio sources could lie behind filaments similar to the one we have discovered. That estimate is only a couple of times larger than the rate of incidence of extreme IHV manifest in the sample of \citet{2003AJ....126.1699L}, suggesting that plasma filaments may be common enough to explain the previously reported examples of extreme scintillation. And in our own data, of course, there are six IHV and five out of the six are behind the filament; the sixth could be behind a different filament, because we don't expect to be able to identify filaments with 100\% reliability (see Section~\ref{subsec:screen}).

It is also notable that the estimated width of our filament ($10^{-3}\;{\rm pc}\simeq 200\;{\rm AU}$) is only about three times larger than has been inferred for the J1819+3845 scattering screen \citep{2015A&A...574A.125D}. The rough  similarity in linear size and in the incidence rate together suggest that a local population of plasma filaments, like the one we have discovered, may be able to explain the IHV/IDV phenomenon as a whole. \tyomaadded{It was previously argued \citep{2013MNRAS.429.2562T} that very similar screens could also be responsible for the parabolic arc phenomenon observed in the secondary spectra of certain radio pulsars \citep{2006ApJ...637..346C, 2004MNRAS.354...43W}. It remains to be seen whether filamentary screens of the kind revealed in this paper can account for other properties of the parabolic arcs, such as their occurrence rate and the scattering anisotropy \citep{2020arXiv200912757R}.}

\subsection{Multiwavelength counterparts}
\label{subsec:multiwavelength}
To clarify the nature of our screen we looked for any possible structure with similar size in multi-wavelength images in Aladin \citep{2000A&AS..143...33B, 2014ASPC..485..277B}. 
No obvious structures were found in H-alpha, CO/HI or continuum images from low frequency radio to gamma-ray.
The lack of an H-alpha detection for our filament is unsurprising given the very low emission measure $\sim10^{-3}\;{\rm cm^{-6}\,pc}$ implied by the size and electron density deduced in the previous section \citep[cf.][]{2020ApJ...900..169M}.

The only structure with similar scale and orientation we noticed is some faint emission in Planck 857 GHz \citep{2016A&A...594A...8P}, which also appears to have a counterpart in GALEX far-ultraviolet images \citep{2007ApJS..173..682M, 2018ApJ...858..101A}. 
Such structures are normally attributed to the material high above the Galactic plane -- molecular hydrogen in dense parts of the Galactic Cirrus clouds and/or dust scattering the UV radiation of the stars in the Galactic disc \citep{1987A&A...183..335J, 2013ApJ...779..180H} -- whereas our plasma filament is local (Section~\ref{subsec:physical_properties}).

Large filamentary structures in the ISM have previously been reported at various wavelengths, including: pulsar bow shocks \citep[e.g.][]{2020ApJ...896L...7D}; Mira-type stellar wind-ISM interactions \citep{2007Natur.448..780M}; and, a possible interstellar shock wave created by an explosion \citep{2020A&A...636L...8B}.\footnote{There are also many filaments seen towards the Galactic Centre \citep{2019Natur.573..235H}. We cannot be sure that they are a fundamentally different phenomenon from our filament; however, the Galactic Centre manifests physical conditions that are very different from the local ISM, and those are non-thermal filaments whereas ours comprises thermal plasma (see Section~\ref{subsec:our_model}).}
We searched for pulsars and stars close to our filament, and found neither nearby pulsars (which are rare towards the Galactic poles), nor nearby stars with direction of proper motion along the same line. 

\begin{figure*}
	\includegraphics[width=2\columnwidth]{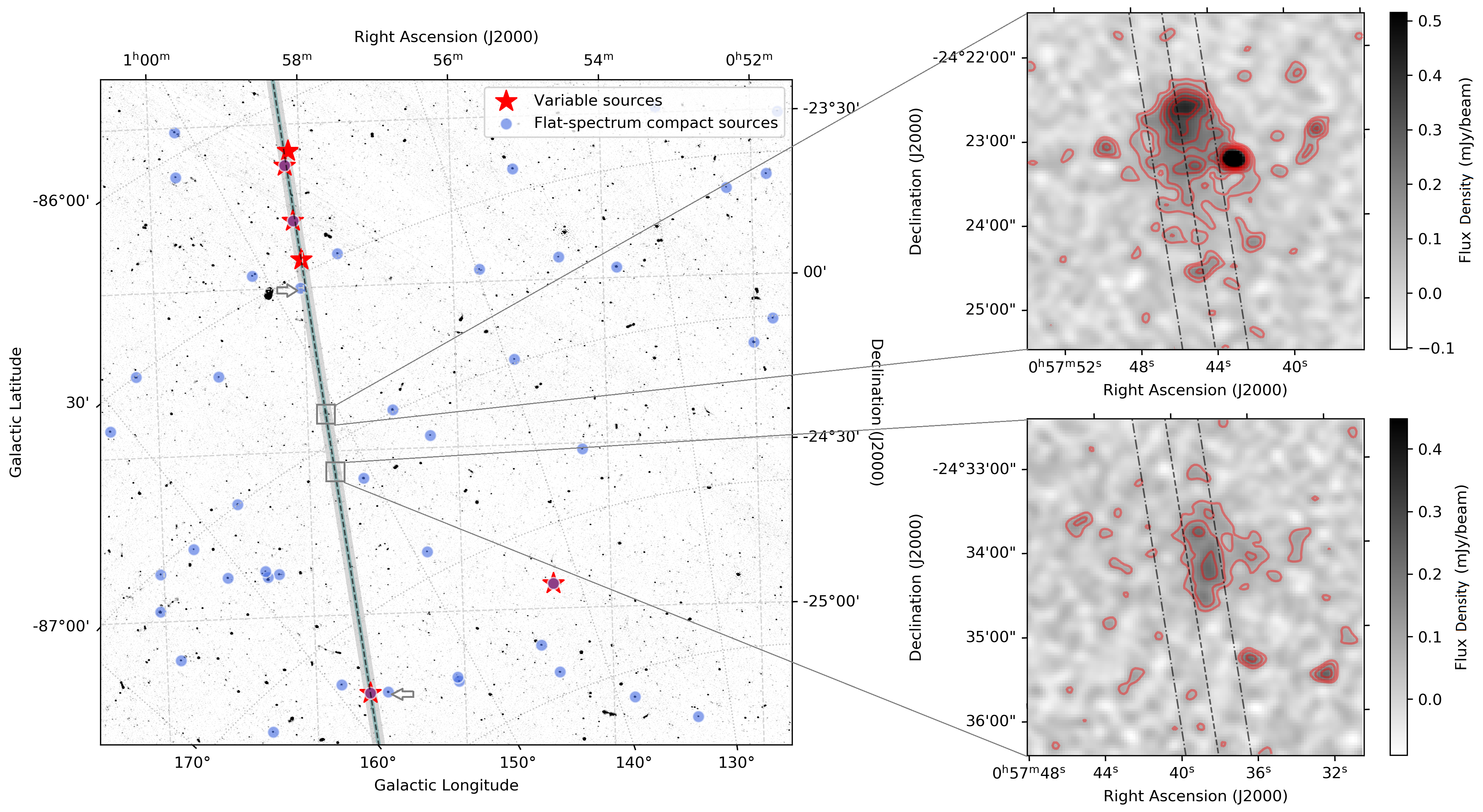}
    \caption{The left panel shows the sky distribution of flat-spectrum, compact sources around our variables. The two arrows point the two non-scintillating sources we used to constrain the upper limit of width of the filament. Right panel shows the two diffuse objects we found in ASKAP images (see discussion in Section~\ref{subsec:multiwavelength}), with separation of $\sim$ 12 arcmin along the line. The dash line represents the best-fit line and the dash-dotted line represents the $\pm 1\sigma$ region. }
    \label{fig:dis_diffuse}
\end{figure*}

We did find two diffuse objects (separation of 12 arcmin) located exactly on the line between the five variables in our ASKAP images (see Fig.~\ref{fig:dis_diffuse}), with morphology similar to the jets of a radio galaxy.
We considered that these might instead be diffuse Galactic emission, and thus potentially associated with our scattering medium. 
We tried to identify a possible host galaxy in optical images, to test the idea that they are lobes of a radio galaxy. 
We found three candidates in DES, and inferred a linear size for the radio sources ranging from 2.1 to 2.8 Mpc, based on photometric redshifts from WISE $\times$ SuperCOSMOS catalogue \citep{2016ApJS..225....5B}. 
Such giant radio galaxies are known, but rare \citep{2017MNRAS.469.2886D}. 
We also checked for emission in low-frequency radio maps, but found no diffuse structure in GaLactic and Extragalactic All-sky Murchison Widefield Array survey \citep[70 -- 231 MHz;][]{2015PASA...32...25W} or TIFR GMRT Sky Survey \citep[150 MHz;][]{2017A&A...598A..78I}. 
We are unable to draw a firm conclusion as to whether these diffuse radio objects that we noticed are Galactic emission or a giant radio galaxy.

\subsection{Interpretation}
We now consider the origins of our filamentary scattering plasma, starting with a comparison between its observed properties and existing ideas about possible sites of enhanced interstellar scattering.

\subsubsection{Current-sheet model}
\citet{2014MNRAS.442.3338P} suggested that interstellar reconnection sheets, aligned with the line-of-sight, could explain key elements of pulsar scintillation phenomenology. In this model the geometry naturally leads to a region of enhanced scattering that is quasi-one-dimensional, around the point(s) where the sheet is tangent to the line-of-sight. However, we observe the region of enhanced scattering to be both straight and narrow (aspect ratio $\sim100{:}1$), and the current-sheet model would require an ad hoc contrivance to reproduce these features. 

The current-sheet model also naturally generates anisotropic scattering, as a result of foreshortening in the plane containing both the line-of-sight and the sheet normal. Our kinematic analysis does point to anisotropic scattering (Section~\ref{subsection:kinematicanalysis}); however, in four out of five cases the observed orientation of the anisotropy is perpendicular to the model prediction. We conclude that the current-sheet model is not a good match to our data.

\subsubsection{Association with hot stars}
For two well-studied IHVs, \citet{2017ApJ...843...15W} established the proximity of the screens to two local A-type stars and proposed that as a paradigm; i.e. extreme scintillation arises in ionised gas associated with hot stars in the solar neighbourhood. In their suggested physical picture the scattering plasma arises as thin skins on tiny molecular gas clouds, with each star carrying a large population of such clouds. That picture has since been shown to be consistent with the kinematics of a third IHV \citep{2019MNRAS.487.4372B}. And the case appears to be further strengthened by the recent discovery of IHV in a source that is surprisingly close to the B-type star Alkaid \citep{2020A&A...641L...4O}. 

However, our data do not match this picture in three respects. Firstly, we are unable to identify a suitable, local, hot star. Secondly, although elongated structures formed part of the proposed picture -- patterned on the ``cometary knots'' of the Helix Nebula -- the aspect ratio of our filament is an order of magnitude larger. Thirdly, we infer anisotropy of the inhomogeneities in the scattering plasma with major axis perpendicular to the long axis of the filament, whereas \citet{2017ApJ...843...15W} imagined those axes to be parallel.

\subsubsection{Association with local absorbing clouds}
\begin{figure}
    \centering
    \includegraphics[width=\columnwidth]{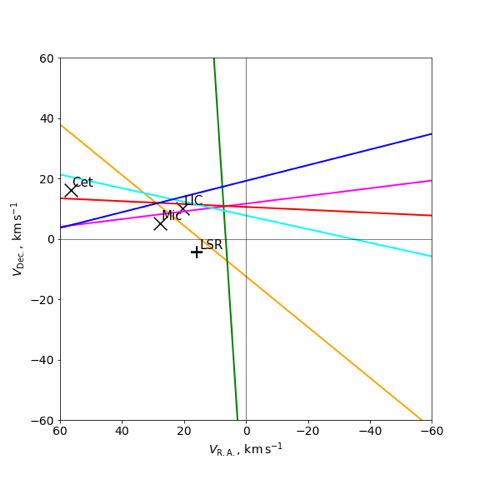}
\caption{Velocity vectors of local warm clouds identified in \citet{2008ApJ...673..283R} projected onto the plane perpendicular to the average celestial position of our six scintillators. The three clouds shown -- LIC, Mic and Cet (marked with crosses) -- are those intersected by this line-of-sight. The position of the local standard of rest (LSR) is also marked. The coloured lines show the best fit one-dimensional kinematic models (Section~\ref{subsection:kinematicanalysis}) for our scintillators; the colour coding is as per Fig.~\ref{fig:rates}.}
    \label{fig:localclouds}
\end{figure}

\citet{2008ApJ...673..283R} identified 15 warm clouds in the local ISM and proposed that interactions at the boundaries of colliding clouds might generate turbulence that would lead to enhanced radio-wave scattering. \citet{2008ApJ...675..413L} found this picture to be consistent with the available kinematic constraints for three well-studied IHV sources. 
As can be seen in Fig.~\ref{fig:localclouds}, the results of our analysis in Section~\ref{subsection:kinematicanalysis} are broadly consistent with the kinematics of the three clouds -- ``Mic'', ``LIC'' and, less favourably, ``Cet'' -- that are foreground to our six variables.  However, our current kinematic constraints, taken in isolation, are not precise enough to allow us to identify a significant match --- as illustrated by the fact that $8/15$ of the \citet{2008ApJ...673..283R} clouds are broadly consistent with our kinematic constraints.

A second point of possible commonality is that two of the foreground clouds (Mic and Cet) were classified as ``filamentary'' by \citet{2008ApJ...673..283R}. However, those clouds are $\sim100^\circ$ in length and $\gtrsim 20^\circ$ in width: huge in comparison with our plasma filament.

The length scales on which electron density fluctuations are required, to explain the scintillation of compact radio sources, is many orders of magnitude smaller than the size of the local absorbing clouds. To generate those fluctuations \citet{2008ApJ...673..283R} and \citet{2008ApJ...675..413L} relied on a turbulent cascade from large scales, with energy input from the relative motions of two clouds at their interface. However, on intermediate scales -- i.e. arcminutes to degrees -- our data reveal an isolated structure that is both straight and thin, quite unlike a snapshot of turbulence. 

\subsubsection{A model inspired by our data}
\label{subsec:our_model}
As none of the foregoing pictures are a good match to our data we are led to consider new ideas, aiming specifically for an interpretation in which the characteristics of our scattering plasma arise naturally. The new insight provided by our observations is the geometry of the scattering screen, so explaining that geometry is our main focus.

There are two key constraints that we can apply to all potential models. First is our estimated number density of screens in the solar neighbourhood (Section~\ref{subsec:physical_properties}) of $\sim 10\;{\rm pc^{-3}}$ --- a few times larger than the density of ordinary stars. Although that estimate is a crude one, based on one detection in one field which is then taken as representative, it is nevertheless a powerful discriminant because it would have to be in error by many orders of magnitude for models based on stellar exotica to be viable. Thus pulsars, runaway stars, giant stars and related phenomena can be ruled out.

The second constraint follows from the fact that the majority of our models have electron densities well above $0.1\,{\rm cm^{-3}}$ (see Fig.~\ref{fig:densitydistance}), where a $\sim 10^4\;{\rm K}$ plasma would typically be in pressure equilibrium with the diffuse ISM \citep{2011ApJ...734...65J}. We reject a picture in which the filament geometry derives from a strong, ordered magnetic field that confines the plasma laterally. The main point against this interpretation is that it would lead to plasma anisotropy with the major axis parallel to the long axis of the filament, contrary to what we observe (Section~\ref{subsection:kinematicanalysis}). Consequently we restrict attention to models which are essentially gas-dynamic, and within that framework an over-pressured plasma should expand laterally at the sound speed, $c_s$. The large aspect ratio of our filament then requires that the source of the plasma is in motion along the length of the filament at a speed $V_*\gtrsim 10^2 c_s$. 

Combining that constraint with a sound speed $c_s\sim 10\;{\rm km\,s^{-1}}$, appropriate to a warm, fully ionised gas, implies $V_*\gtrsim 10^3\;{\rm km\,s^{-1}}$. That speed is problematic because it is much larger than the speeds of almost everything in the solar neighbourhood -- e.g. low-mass dwarf stars have a velocity dispersion of $\simeq 30\;{\rm km\,s^{-1}}$ -- and even exceeds the escape speed from the Galaxy. In response to this problem we turn to a picture in which the sound speed is much lower, with $c_s\lesssim 0.3\;{\rm km\,s^{-1}}$ so that $V_*\lesssim 30\;{\rm km\,s^{-1}}$ yields an acceptable aspect ratio. Such a low sound speed implies a low temperature ($\lesssim 10\;{\rm K}$ for hydrogen), and is only plausible if the plasma is a trace component within a gas that is predominantly neutral.

We are thus led to a picture in which the underlying astrophysical phenomenon is a directed stream of cold gas. Such flows are not a feature of our current description of either the ISM or of the mass-loss from main-sequence stars, making broad swathes of possible models look immediately unattractive. We have also checked the Gaia \citep{2018A&A...616A...1G} and Hipparcos \citep{2007A&A...474..653V} catalogues for nearby stars whose motion lies within the plane of our filament, but we found nothing surprising. 

We are aware of only one published prediction of a phenomenon that resembles what we require: in the context of galactic nuclei, \citet{2016ApJ...822...48G} pointed out that the tidal disruption of stars by massive black holes should lead to thin streams of unbound gas. As it stands that model does not apply to the solar neighbourhood. However, based on  modelling of the internal structure of hydrogen snow clouds, \citet{2019ApJ...881...69W} pointed out that such clouds would be tidally disrupted by stars. Thus if similar clouds are abundant in the solar neighbourhood -- as has been considered by many authors \citep[e.g.][]{1994A&A...285...79P,1996ApJ...472...34G,1998ApJ...498L.125W,2017ApJ...843...15W} -- then we expect that tidal disruptions will be frequent, and tidal streams of cold gas will be common.

In a tidal stream interpretation of our plasma filament there is a natural source of density fluctuations on a range of scales: at the boundary between the stream \tyomaadded{launched by the tidal disruption} and the ambient ISM there is a strong velocity shear so the interface will be Kelvin-Helmholtz unstable, leading ultimately to turbulence.
Although the hypothesis of a neutral gas stream does not in itself guarantee the presence of plasma, which is needed to generate scintillations, there are several possible sources of ionisation -- cosmic rays, photoionisation of metals, and shock heating in the case of high stream velocities -- so generating a small ionised fraction should not be troublesome. Finally we note that, although our estimates suggest that the plasma is probably over-pressured relative to the ambient ISM, a tidal stream interpretation does not demand it and all of the solutions in Fig.~\ref{fig:densitydistance} can be accommodated.

\section{Conclusions}
\label{sec:conclusions}

We conducted an unbiased search for highly variable sources on timescales of hours with ASKAP, detecting six rapid scintillators among $\sim$ 40\,000 sources in a 30 deg$^2$ field. 
Our variables include two sources showing modulation indices of up to $\sim 40\%$, which are new examples of the rare, extreme IHV phenomenon. 
A surprising discovery is the existence of a degree-long plasma filament, revealed by five scintillators in a line on the sky. 
We constrained the geometric boundary of the screen, for the first time, and obtained a length greater than 1.7 degree and width between $1$ and $4$ arcmin. \tyomaadded{We note that the Sourthern-most source along the filament is well separated from the four other sources along the filament which extend over only $\sim20\,\mathrm{arcmin}$ along its length. Had it happenned to lie there by chance the degree to which the filament is narrow and straight is diminished somewhat. The probability of the variable being found along the continuation of the four-sources implied filament out of all other places in the field is low (less than one per cent) but this might not be a sufficient argument against chance alignment given the singularity of the case. However, the similarity of the scattering properties of this variable to those inferred for the other four suggests strongly that they all lie behind the same narrow, long and remarkably straight physical structure.}

We measured the annual modulation of the scintillation rate and found that the plasma microstructure is highly anisotropic, with major axis roughly perpendicular to the long axis of the filament. 
These properties do not accord with any published suggestions for the origin of extreme scattering. Instead we propose a picture in which the plasma is a trace component within a cold, neutral gas stream, and we interpret that stream as a tidal remnant.
The interpretation mainly comes from analysis of the five aligned sources, and we will give further consideration of the sixth variable source (the one not in the line) in a subsequent paper. 
Irrespective of the origin of the plasma filament that we observe, the size and likely sky-covering fraction of similar filaments suggest that they can probably account for the IHV/IDV phenomenon as a whole.

This is the first time that multiple scintillators have been detected behind the same plasma screen, demonstrating the power of ASKAP's combination of a large field-of-view with high sensitivity.
Using similar imaging and search techniques we expect to detect a large sample of similar variables in future ASKAP sky surveys; that sample will yield detailed information on the discrete plasma structures in the solar neighbourhood. 

\section*{Acknowledgements}
We thank Jim Cordes, Lister Staveley-Smith, Elaine Sadler, Ron Ekers, David McConnell, Lucyna Kedziora-Chudczer, Phil Edwards \tyomaadded{and Bill Coles} for useful suggestions. We also thank CSIRO staff Aidan Hotan and Vanessa Moss for scheduling additional ASKAP observations.
\newadded{We thank the anonymous referee for the extremely helpful comments.}
YW is supported by the China Scholarship Council. 
TM acknowledges the support of the Australian Research Council through grants FT150100099 and DP190100561. 
DK is supported by NSF grant  AST-1816492. 
This work used resources of China SKA Regional Centre prototype \citep{2019NatAs...3.1030A} funded by the National Key R\&D Programme of China (2018YFA0404603) and Chinese Academy of Sciences (114231KYSB20170003). 
This research has made use of the VizieR catalogue access tool (DOI: 10.26093/cds/vizier, \citealt{2000A&AS..143...33B}) and SIMBAD \citep{2000A&AS..143....9W}, operated by CDS, Strasbourg, France; Astropy,\footnote{http://www.astropy.org} a community-developed core Python package for Astronomy \citep{astropy:2013, astropy:2018}, {\tt numpy} \citep{harris2020array}, {\tt scipy} \citep{2020SciPy-NMeth}, {\tt matplotlib} \citep{matplotlib}, {\tt emcee} \citep{2013PASP..125..306F}, {\tt celerite} \citep{celerite} and {\tt scintools} \citep{2020arXiv200912757R}. 
The Australian Square Kilometre Array Pathfinder is part of the Australia Telescope National Facility which is managed by CSIRO. 
Operation of ASKAP is funded by the Australian Government with support from the National Collaborative Research Infrastructure Strategy. ASKAP uses the resources of the Pawsey Supercomputing Centre. Establishment of ASKAP, the Murchison Radio-astronomy Observatory and the Pawsey Supercomputing Centre are initiatives of the Australian Government, with support from the Government of Western Australia and the Science and Industry Endowment Fund. 
We acknowledge the Wajarri Yamatji as the traditional owners of the Murchison Radio-astronomy Observatory site. 

\section*{Data availability}

The ASKAP data (SB9602, SB9649, SB9910, SB10463, SB12704, SB13570, and SB15191) used in this paper can be accessed through the CSIRO ASKAP Science Data Archive (CASDA).





\bibliographystyle{mnras}
\bibliography{example} 




\appendix

\section{Details of variable sources}
\label{sec:details_variables}

\begin{figure*}
	\includegraphics[width=2\columnwidth]{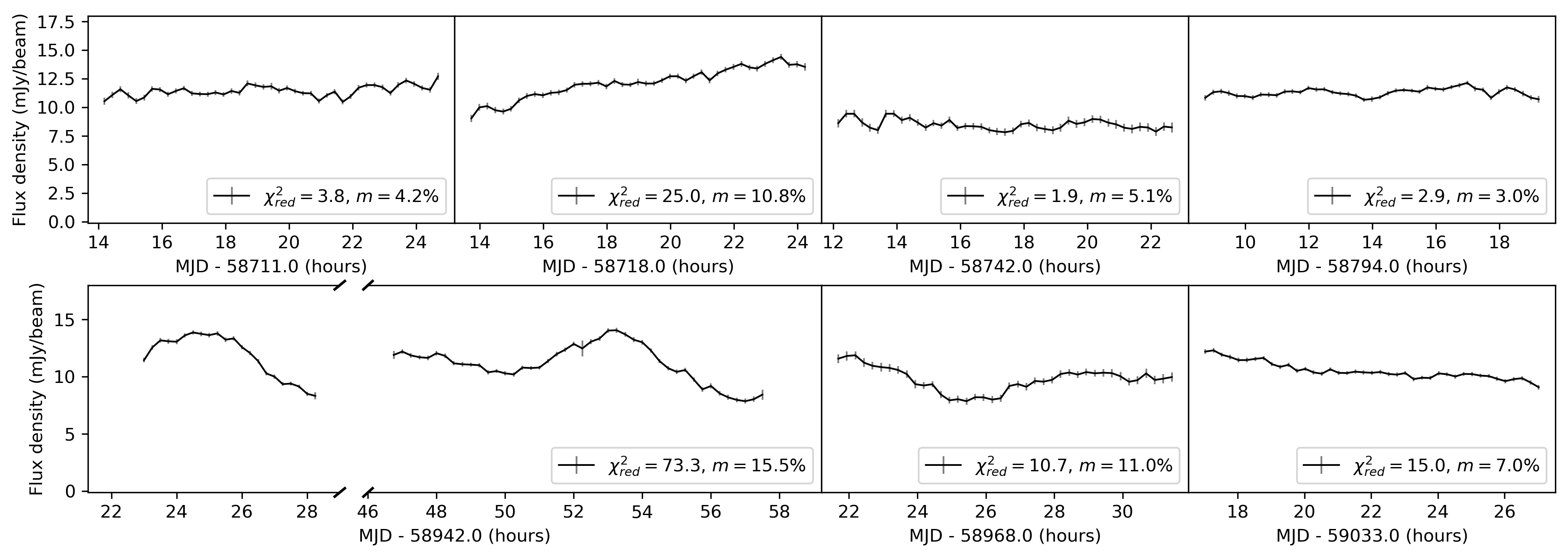}
    \caption{Light-curves of source J005806.72--234744.63. Details as in Fig.~\ref{fig:lc_var_J005800}. }
    \label{fig:lc_var_J005806}
\end{figure*}

\begin{figure*}
	\includegraphics[width=2\columnwidth]{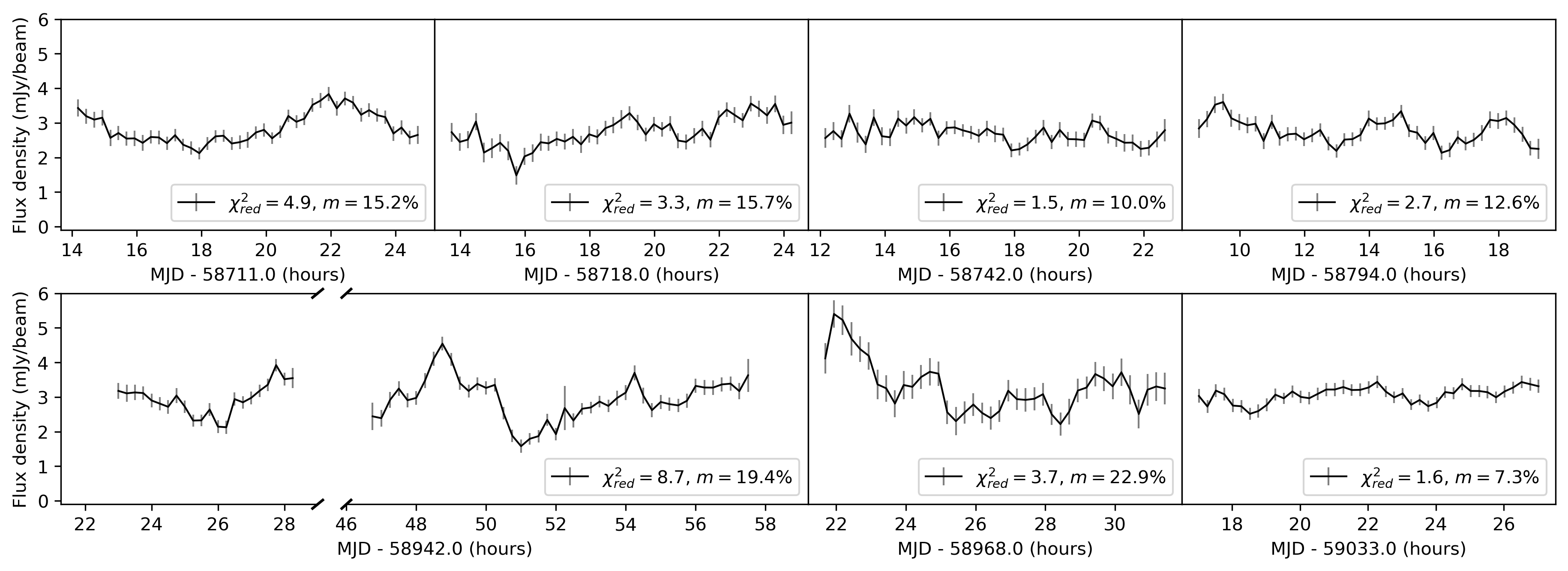}
    \caption{Light-curves of source J005809.00--233454.00. Details as in Fig.~\ref{fig:lc_var_J005800}. }
    \label{fig:lc_var_J005809}
\end{figure*}

\begin{figure*}
	\includegraphics[width=2\columnwidth]{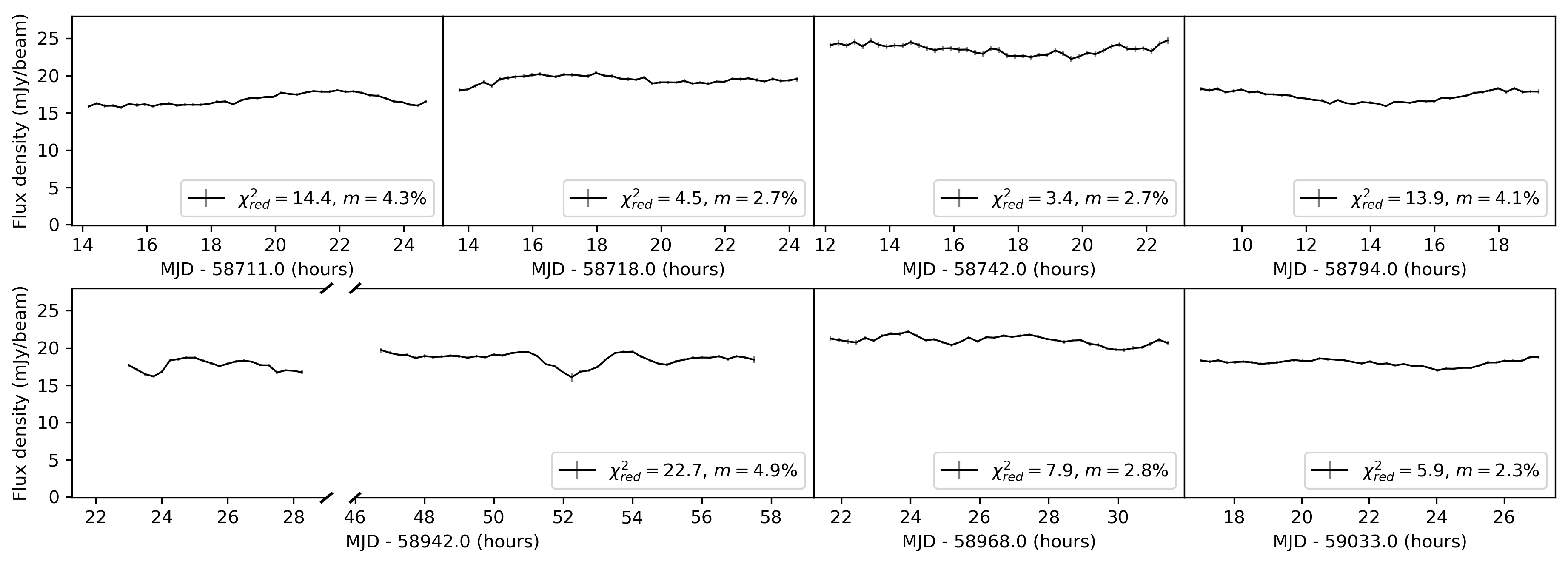}
    \caption{Light-curves of source J005716.91--251424.64. Details as in Fig.~\ref{fig:lc_var_J005800}. }
    \label{fig:lc_var_J005716}
\end{figure*}

\begin{figure*}
	\includegraphics[width=2\columnwidth]{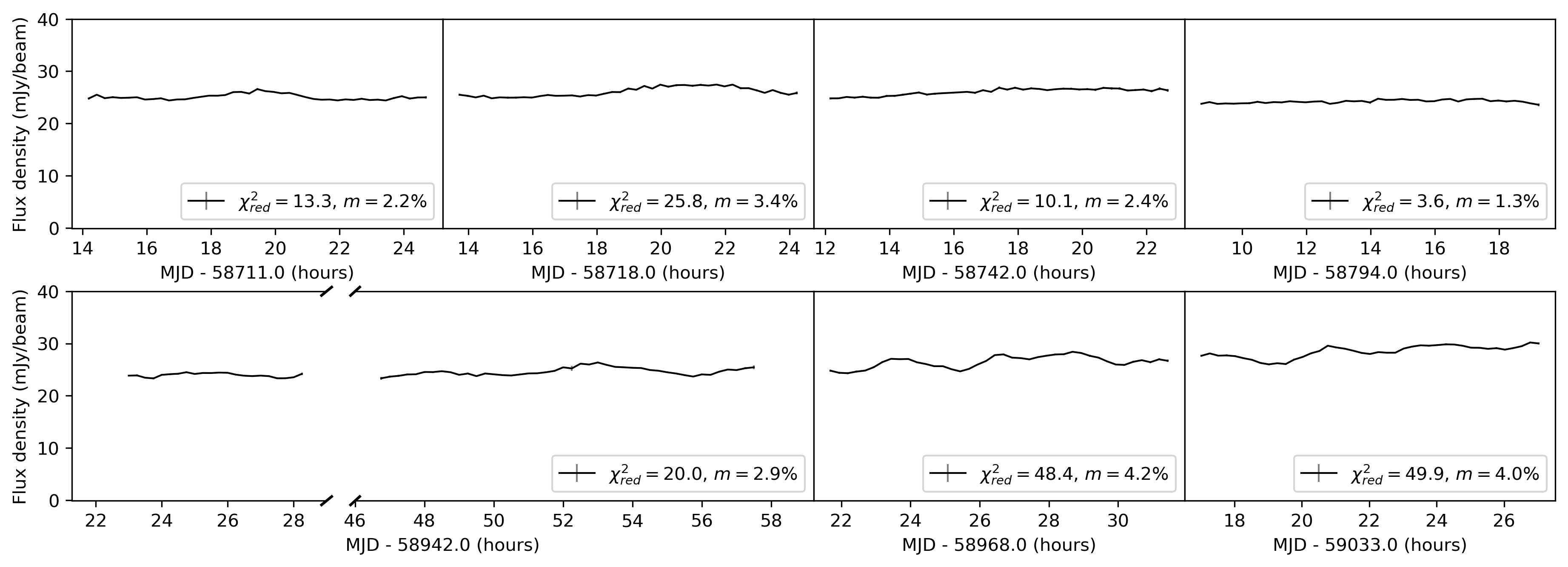}
    \caption{Light-curves of source J005446.77--245529.30. Details as in Fig.~\ref{fig:lc_var_J005800}. }
    \label{fig:lc_var_J005446}
\end{figure*}

\begin{figure*}
	\includegraphics[width=2\columnwidth]{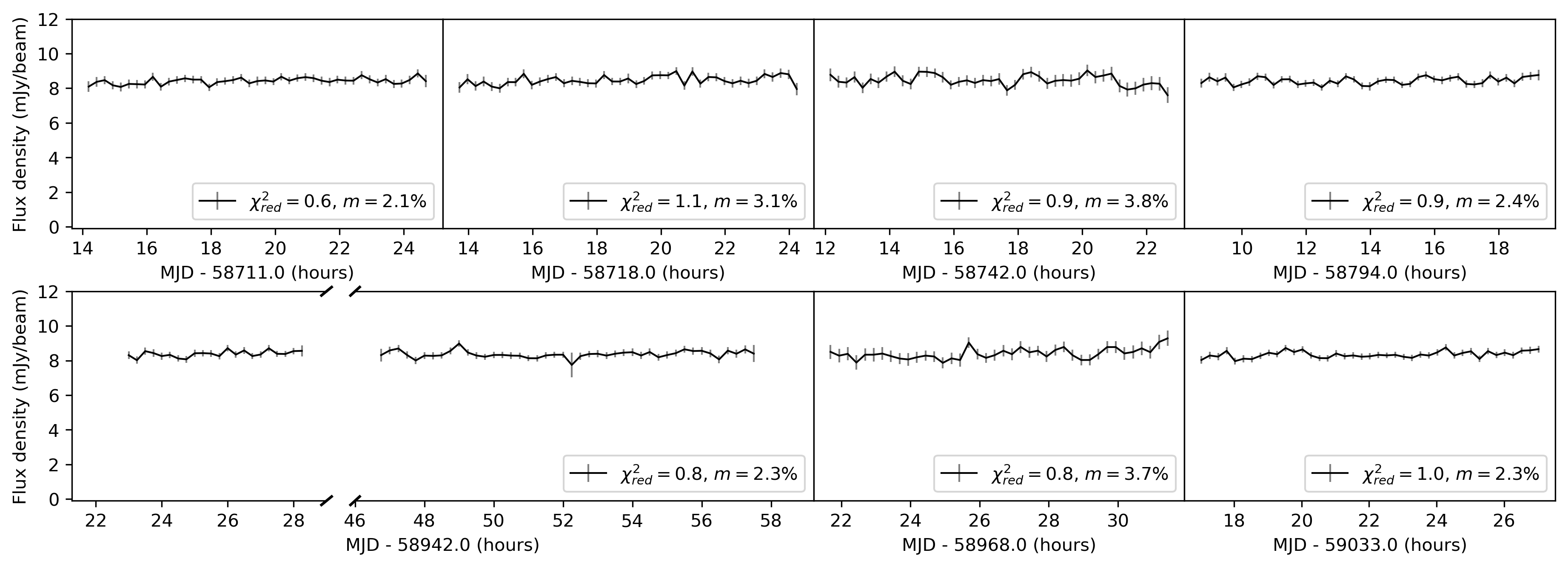}
    \caption{Light-curves of reference source J005806.62--234306.98, a non-variable source also near the line with offset of $\sim13''$. Details as in Fig.~\ref{fig:lc_var_J005800}. }
    \label{fig:lc_ref_J005806}
\end{figure*}

This appendix contains light-curves of the other four variable sources in all epochs, see Fig.~\ref{fig:lc_var_J005806} for source J005806.74--234744.63, Fig.~\ref{fig:lc_var_J005809} for source J005809.00--233454.00, Fig.~\ref{fig:lc_var_J005716} for source J005716.91--251424.64, and Fig.~\ref{fig:lc_var_J005446} for J005446.77--245529.30. 
We also included the light-curves of the reference source J005806.62-234306.98 (see Fig.~\ref{fig:lc_ref_J005806}), which is a non-variable source close to the best-fitted projection line. 
Fig.~\ref{fig:slices} shows movies of two extreme scintillators J005800.94--235449.11 and J005812.02--233735.39. 
\newadded{Each movie contains 43 frames, made by a sequence of contiguous 15-min model-subtracted images of epoch 1. }

\begin{figure*}
    \animategraphics[controls,width=\columnwidth]{25}{Figures/slices/J005800_}{0}{42}
    \animategraphics[controls,width=\columnwidth]{25}{Figures/slices/J005812_}{0}{42}
    \caption{The animation of 15-minute model-subtracted images (worked in Adobe reader). The left panel is for the source J005800.94--235449.11 (centred on the image) in epoch 1, and the right panel is for the source J005812.02--233735.39 in epoch 1. \newadded{Each movie contains a series of 43 images. The upper right shows the flux density and the rms noise measured at the source position (marked as plus), followed by the sequence number. }}
    \label{fig:slices}
\end{figure*}

\bsp	
\label{lastpage}
\end{document}